\documentclass[usegraphicx,useAMS,usenatbib]{mn2e}
\usepackage{graphicx}
\usepackage{rotating}
\usepackage{subfigure}

\title[No age spread in the ONC]
 {No wide spread of stellar ages in the Orion Nebula Cluster}

\author[R.D. Jeffries et al.]
  {R.D.~Jeffries$^1$, S. P. Littlefair$^2$, Tim Naylor$^3$, N. J. Mayne$^3$\\
  $^1$ Astrophysics Group, Research Institute for the Environment, Physical
  Sciences and Applied Mathematics,\\ Keele University, Staffordshire ST5
  5BG, UK\\
  $^2$ Department of Physics and Astronomy, University of Sheffield, S3
  7RH, UK\\
  $^3$ School of Physics, University of Exeter, Exeter, EX4 4QL, UK}
\date{Submitted 2 May 2011}

\pagerange{\pageref{firstpage}--\pageref{lastpage}} \pubyear{2010}

\def\LaTeX{L\kern-.36em\raise.3ex\hbox{a}\kern-.15em
    T\kern-.1667em\lower.7ex\hbox{E}\kern-.125emX}

\begin{document}
\label{firstpage}
\maketitle

\begin{abstract}
The wide luminosity dispersion seen for stars at a given effective
temperature in the Hertzsprung-Russell diagrams of young clusters and
star forming regions is often interpreted as due to significant ($\sim
10$\,Myr) spreads in stellar contraction age. In the scenario where
most stars are born with circumstellar discs, and that disc signatures
decay monotonically (on average) over timescales of only a few Myr,
then any such age spread should lead to clear differences in the age
distributions of stars with and without discs.  We have investigated
large samples of stars in the Orion Nebula Cluster (ONC) using three
methods to diagnose disc presence from infrared measurements. We find
no significant difference in the mean ages or age distributions of
stars with and without discs, consistent with expectations for a coeval
population. Using a simple quantitative model we show that any real age
spread must be smaller than the median disc lifetime. For a log-normal
age distribution, there is an upper limit of $<0.14$\,dex (at 99 per
cent confidence) to any real age dispersion, compared to the $\simeq
0.4$ dex implied by the Hertzsprung-Russell diagram. If the mean age of
the ONC is 2.5\,Myr, this would mean at least 95\% of its low-mass
stellar population has ages between 1.3--4.8\,Myr.  We suggest that the
observed luminosity dispersion is caused by a combination of
observational uncertainties and physical mechanisms that disorder the
conventional relationship between luminosity and age for pre
main-sequence stars.  This means that individual stellar ages from the
Hertzsprung-Russell diagram are unreliable and cannot be used to
directly infer a star formation history. Irrespective of what causes
the wide luminosity dispersion, the finding that any real age
dispersion is less than the median disc lifetime argues strongly
against star formation scenarios for the ONC lasting longer than a few
Myr.
\end{abstract}


\section{Introduction}
When stars are newly born and emerge from their natal clouds onto the
pre main-sequence (PMS), they can be placed in the Hertzsprung-Russell
(HR) diagram. Low-mass stars ($\leq 2\,M_{\odot}$) take $>10$\,Myr to
descend their Hayashi tracks, commence hydrogen burning and settle onto
the zero age main sequence. During this time the luminosity and
effective temperature ($T_{\rm eff}$) of a PMS star can, in principle,
be used to determine a unique, although model-dependent, age.

This technique has been used in many young clusters and star forming
regions (SFRs) as the basis for inferring the age distribution and
hence the star formation history. Examples range widely;
from nearby, comparatively sparse clusters \citep{palla99}, to
very rich and dense clusters within our own Galaxy \citep{beccari10},
or even individual clusters in the Magellanic clouds \citep{dario10a}.  
A significant luminosity spread,
of an order of magnitude or more at a given $T_{\rm eff}$, is almost
invariably found and this has been used to infer age spreads of order
10\,Myr within a cluster or SFR \citep{palla00}.

Others dispute the reality of such large age spreads and debate whether
the effects of unresolved binarity, differential extinction, the
contribution of accretion luminosity, photometric variability,
varying accretion histories over millions of years and
observational uncertainties in spectral types and magnitudes could all
contribute to the observed luminosity dispersion to some extent
\citep{hartmann01, burningham05}. \citet{hillenbrand08} showed that it
may be difficult to verify or quantify age spreads unless the
observational uncertainties are small and the size and distribution of
the astrophysical sources of scatter are well understood. Jeffries
(2007) showed that for the Orion Nebula Cluster (ONC -- see below),
there is a spread of radius at a given $T_{\rm eff}$, consistent with
the luminosity spread observed. However, this paper also cautions that
a spread of radius may not imply a spread in age. For example, some
models suggest that co-eval stars with differing early
accretion histories can still have significantly different radii several
million years later (Tout, Livio \& Bonnell 1999; 
Baraffe, Chabrier \& Gallardo 2009).

Resolving this issue is important because ages from the HR diagrams of
young SFRs are used to investigate different star formation scenarios,
calculate star formation rates and set the clock for the dispersal of
circumstellar material and the formation of planetary systems. For
example, the inference of a large age spread in SFRs has been taken (by
some) as evidence against a ``fast'' mode of star formation
governed by the rapid dissipation of supersonic turbulence
\citep{elmegreen00, hartmannsfr01, vazquez05}, and used instead to support 
a ``slow'' mode of star formation, where collapse
is regulated by the ambipolar diffusion of strong magnetic fields
\citep{tassis04, tan06}. 

The ONC is one of the best-studied nearby SFRs \citep{jones88,
hillenbrand97}.  \citet{palla99} used the HR diagram to deduce that
star formation in the ONC began at least 10\,Myr ago and has
accelerated up to the present day. \citet{huff06} extended this work to
show that the inferred star formation history is similar in both the
outer and inner parts (the Trapezium) of the ONC and for stars of all
masses. These authors hypothesize that the ONC formed from a cloud
supported by mildly dissipative turbulence, which collapsed globally in
a quasi-static, but accelerating fashion over 10\,Myr.  Contrary to
this, \citet{furesz08} and \citet{tobin09} argue that the close
agreement between the kinematics of the stars and molecular gas in the
ONC implies that the ONC is younger than a crossing time ($\leq
1$\,Myr). The ONC HR diagram has recently been updated and improved by
\citet{dario10b} (hereafter referred to as DR10), who present revised
determinations of luminosities and $T_{\rm eff}$ and derive the
distribution of members in the mass-age plane using evolutionary
models. They found a mean age of 2--3\,Myr, but with a very similar
dispersion in luminosity, and hence inferred age spreads, to the
earlier studies.

If the ONC is very young and undergoing rapid collapse, that would be
difficult to reconcile with the age spread and mean age found on the
basis of the HR diagram by \citet{palla99} and DR10. In this paper the
reality of the ONC age spread is re-examined using the crude, but
independent clock afforded by the time-dependent dispersal of
circumstellar material around young PMS stars. In Section 2 the methods
and the observational material are explained. The results are presented
in Section 3 and examined with a simple interpretive model in Section
4. Section 5 discusses the results and their implications for PMS age
determinations. Section 6 provides a summary.

\section{Methods and Observational Data}
\subsection{Disc dispersal as an independent clock}

\begin{figure*}
\centering
\begin{minipage}[t]{0.33\textwidth}
\includegraphics[width=55mm]{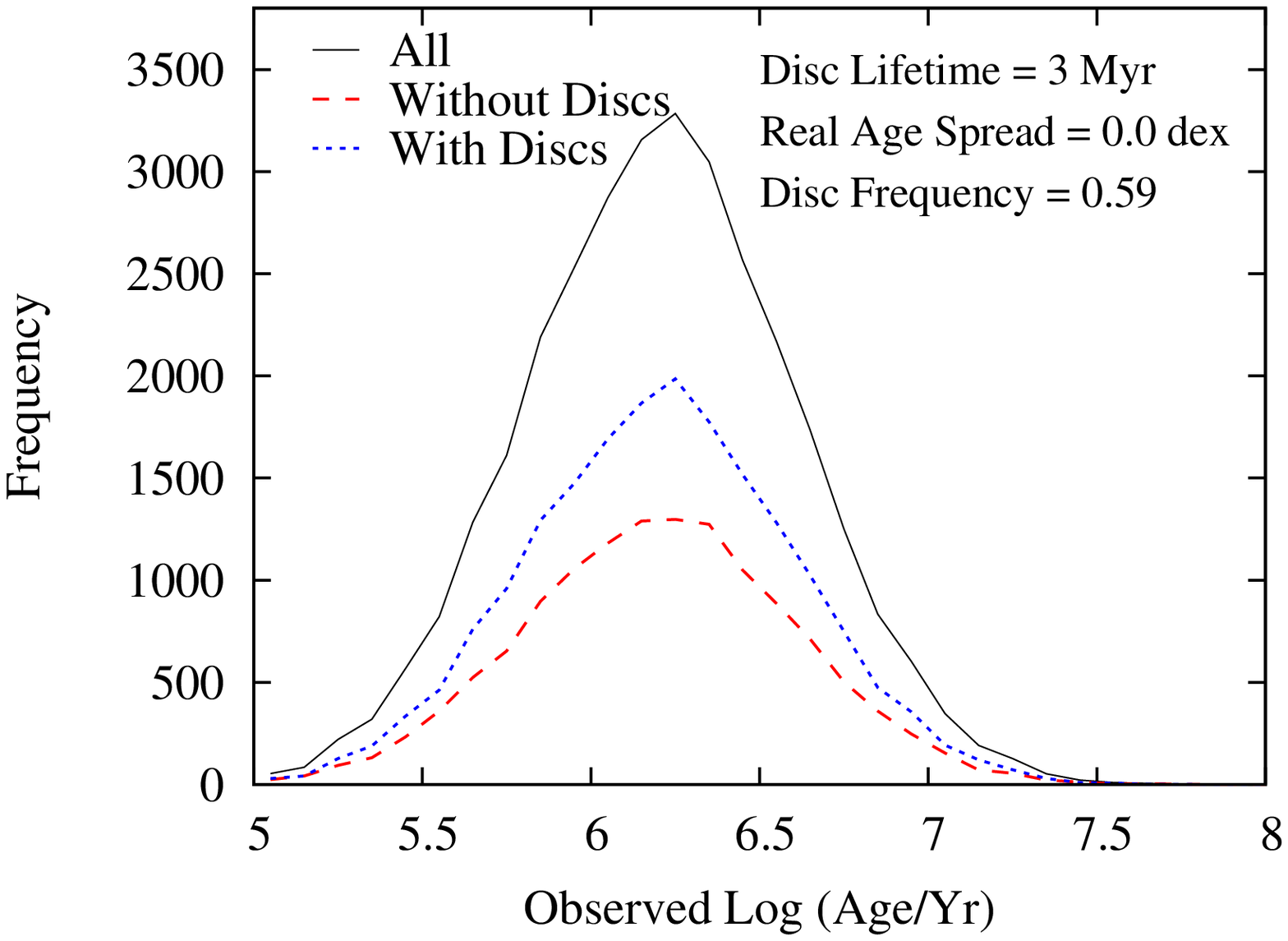}
\end{minipage}
\begin{minipage}[t]{0.33\textwidth}
\includegraphics[width=55mm]{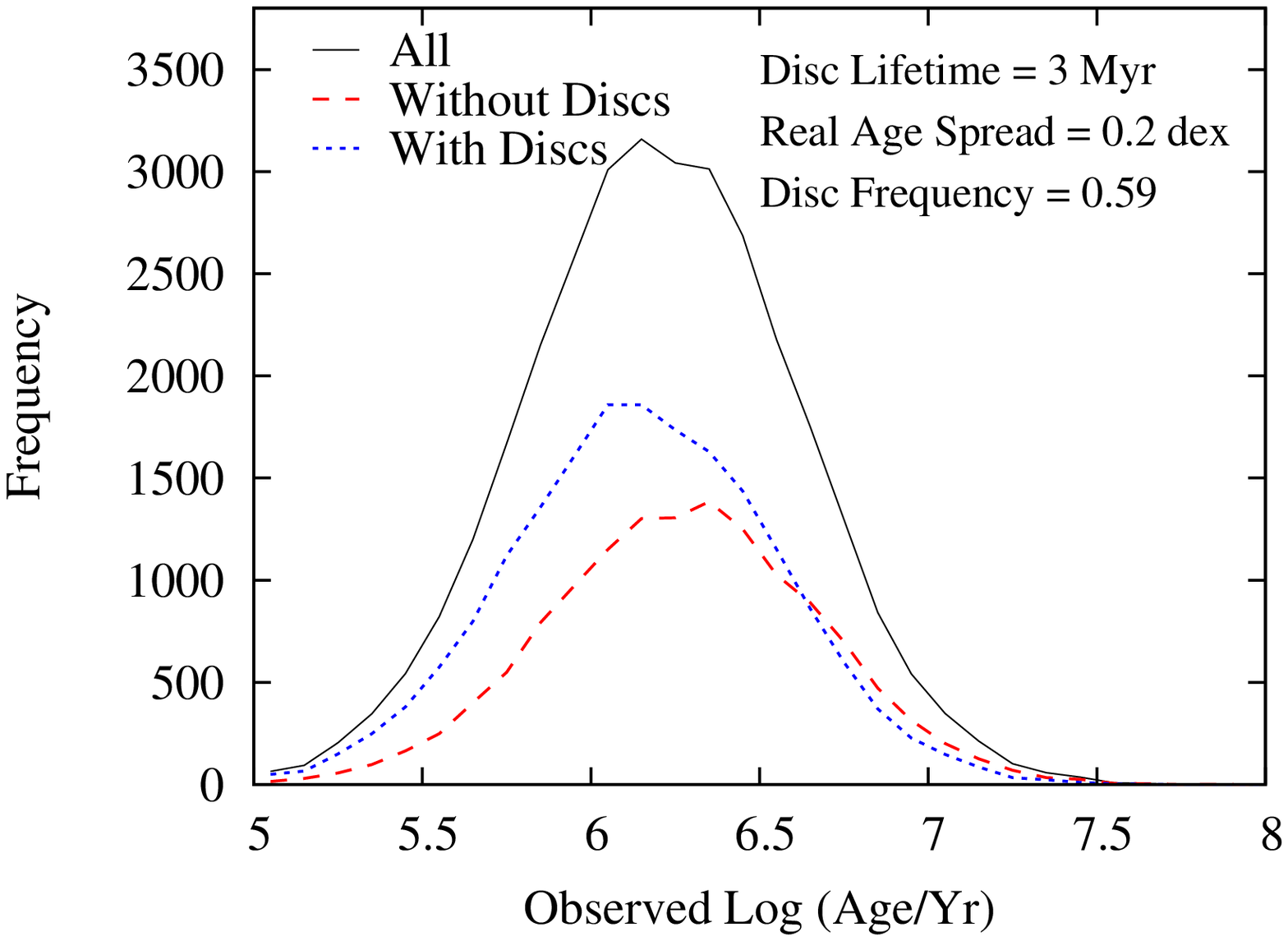}
\end{minipage}
\begin{minipage}[t]{0.33\textwidth}
\includegraphics[width=55mm]{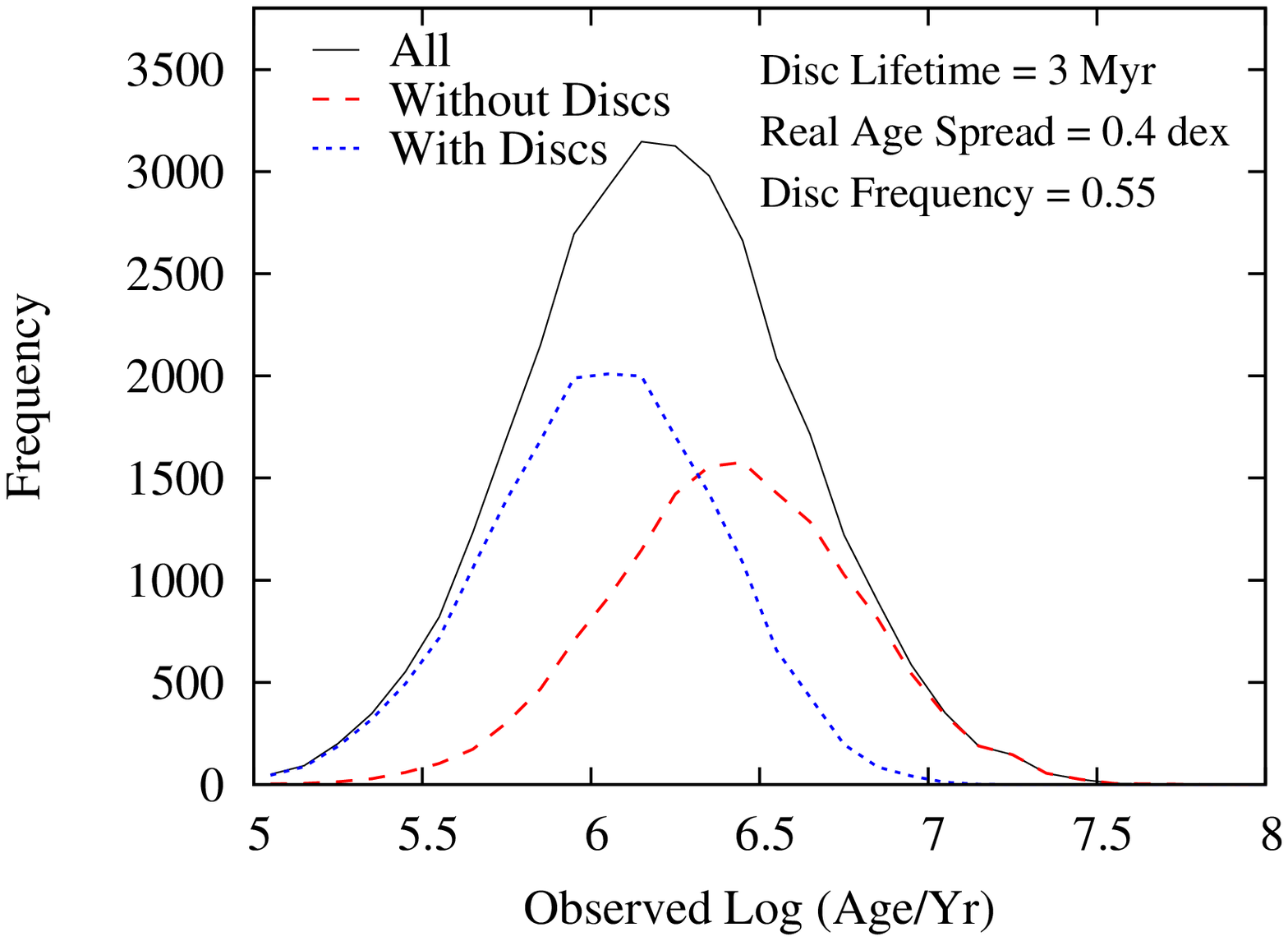}
\end{minipage}

\caption{The apparent age distributions of stars with and without discs 
  generated for simulated
  populations of 30\,000 stars in a cluster with a mean $\log$ age (in years)
  of 6.2, an observed dispersion in $\log$ age of 0.4\,dex and an exponential
  disc decay lifetime of 3\,Myr. The three panels show cases where the
  real dispersion of age in the stellar population (as opposed to
  dispersion caused by observational uncertainties etc.) is increased
  from zero (a coeval population) to 0.2\,dex (a mix of effects) to 0.4\,dex (where the entire
  observed age dispersion is due to a real age dispersion). These
  plots show the differences expected in the age distributions
  of stars with and without discs once there are
  significant numbers of stars that are older than the mean disc lifetime.}
\label{showmodel}
\end{figure*}

It is well known that at the earliest ages, most, if not all, PMS stars
are surrounded by optically thick circumstellar discs. These can be
revealed either by the infrared flux emitted by warm dust in the disc,
or the signatures of gas accretion from the disc onto the star.  Groups
of stars in young clusters and SFRs can be used to determine the
timescale for the dispersal of inner disc material, traced by
near-infrared excesses. Plotting the fraction of stars with a $K$-band
excess versus the mean age (deduced from the HR diagram) of clusters,
suggests that half of PMS stars lose their inner discs in about 3\,Myr
and that the timescale for almost all stars to lose their discs is
about 6\,Myr \citep{hillenbrand05}.  Observations at these relatively
short wavelengths may not produce a complete census of discs \citep[see
the discussion in][]{lada00}, however whilst \citet{haisch01} found
somewhat higher disc frequencies using $K-L$ excess as a disc
indicator, the derived disc dispersal timescale was similar.

This work has now been expanded extensively using more sensitive
Spitzer data, but with little change in the overall conclusion. There
is a good correlation between $K-L$ excesses at wavelengths of
2.0--3.5\,$\mu$m and excesses at the longer 3.6--8.0\,$\mu$m
wavelengths probed by Spitzer. Additional data from more clusters
\citep[e.g. see][]{hernandez08, kennedy09} has strengthened the
conclusion that the fraction of stars with primordial discs declines
with age, such that most discs have dispersed after 5\,Myr,
although a few per cent of stars maintain some circumstellar material
in clusters with an age of $\simeq 10$\,Myr.

In principle the declining disc fraction with age seen in an ensemble
of clusters could be used as a means of constraining any age spread
within a single cluster.  However, it is not clear to what extent the
fraction of stars with discs in a single cluster is determined by the
disc lifetime or a spread of ages within that cluster. A disc fraction
that decreases with mean cluster age could be produced by a range of
disc lifetimes in increasingly elderly but strictly coeval cluster
populations. On the other hand, the trend in disc fraction could also
be explained if the cluster populations had a spread of ages, with
clusters of increasing mean age possessing larger proportions of stars
older than some unique disc lifetime. These two possibilities would
have different signatures in the present-day HR diagram and in the
comparative age distributions of stars with and without discs.  For the
first possibility we would expect to see no luminosity spread in the HR
diagram beyond that contributed by astrophysical scatter (binarity,
variability etc.) and observational uncertainties, and no correlation
between the presence of a disc and HR diagram position. However, for
the second possibility we would expect a clear distinction in the HR
diagram and inferred age distributions between young stars with discs
and older stars that had lost their discs. For a more general case
between these two extremes (i.e. an age spread {\it and} a range of
disc lifetimes) we would expect to see a strong correlation between age
determined from the HR diagram and the presence of a disc whenever the
age spread becomes comparable to, or exceeds the mean disc lifetime.

To illustrate this argument, Fig.~\ref{showmodel} shows a simulation
using a model that is explored in more detail in Section~4. It is
assumed (for the purposes of demonstration) that the observed
distribution of ages from the HR diagram can be represented as
log-normal in age (a reasonable approximation for the ONC
discussed further in Section~4) with a mean log age of 6.2 (in years),
a dispersion $\sigma = 0.4$~dex, and that this dispersion is formed
from the quadrature sum of observational uncertainties, binarity,
variability (etc.) and a separate contribution due to a {\it real}
dispersion in age. It is further assumed that all stars begin their
lives with a disc and that the disc lifetime is drawn from an
exponentially decaying distribution with a characteristic timescale of
3\,Myr. Different functional forms for these distributions are possible
and will be explored in Section~4.  The three panels of
Fig.~\ref{showmodel} show, for simulated populations of 30\,000 stars,
how the observed (or apparent) age distributions for stars with and without discs
become clearly separated when the dispersion in real ages becomes large
enough that there are many stars older than 3\,Myr that have a high
probability of having lost their discs.

There have been a number of previous attempts to identify this
phenomenon with mixed outcomes. \citet{strom89} found that there was
considerable overlap of stars with and without near-infrared excess in
the HR diagram of the Taurus-Auriga association, but that the stars
presumed to be discless were older on average. Subsequently,
\citet{hartigan95} and \citet{bertout07} found that classical T-Tauri
stars (CTTS) with veiling and accretion discs were systematically
younger than their discless, weak-lined T-Tauri star (WTTS)
counterparts in the Taurus-Auriga association. These samples were
relatively small and it is likely that X-ray selected foreground field
stars contaminated the WTTS samples, making them appear older.
\citet{fang09} observed CTTS and WTTS in the L1630 and L1641 clouds,
finding some evidence for a decrease with age in disc frequency
determined from infrared excesses, and a 97 per cent signficance result
that CTTS and WTTS were not drawn from the same age distribution. On
the other hand \citet{herbig98}, \citet{herbig02}, \citet{dahm05} and
\citet{rigliaco11} all found no difference in the age distributions of
CTTS and WTTS in the IC\,348, IC\,5146, NGC~2264 and $\sigma$~Orionis
clusters respectively.

\subsection{Observations of the ONC}

\begin{figure}
\centering
\includegraphics[width=80mm]{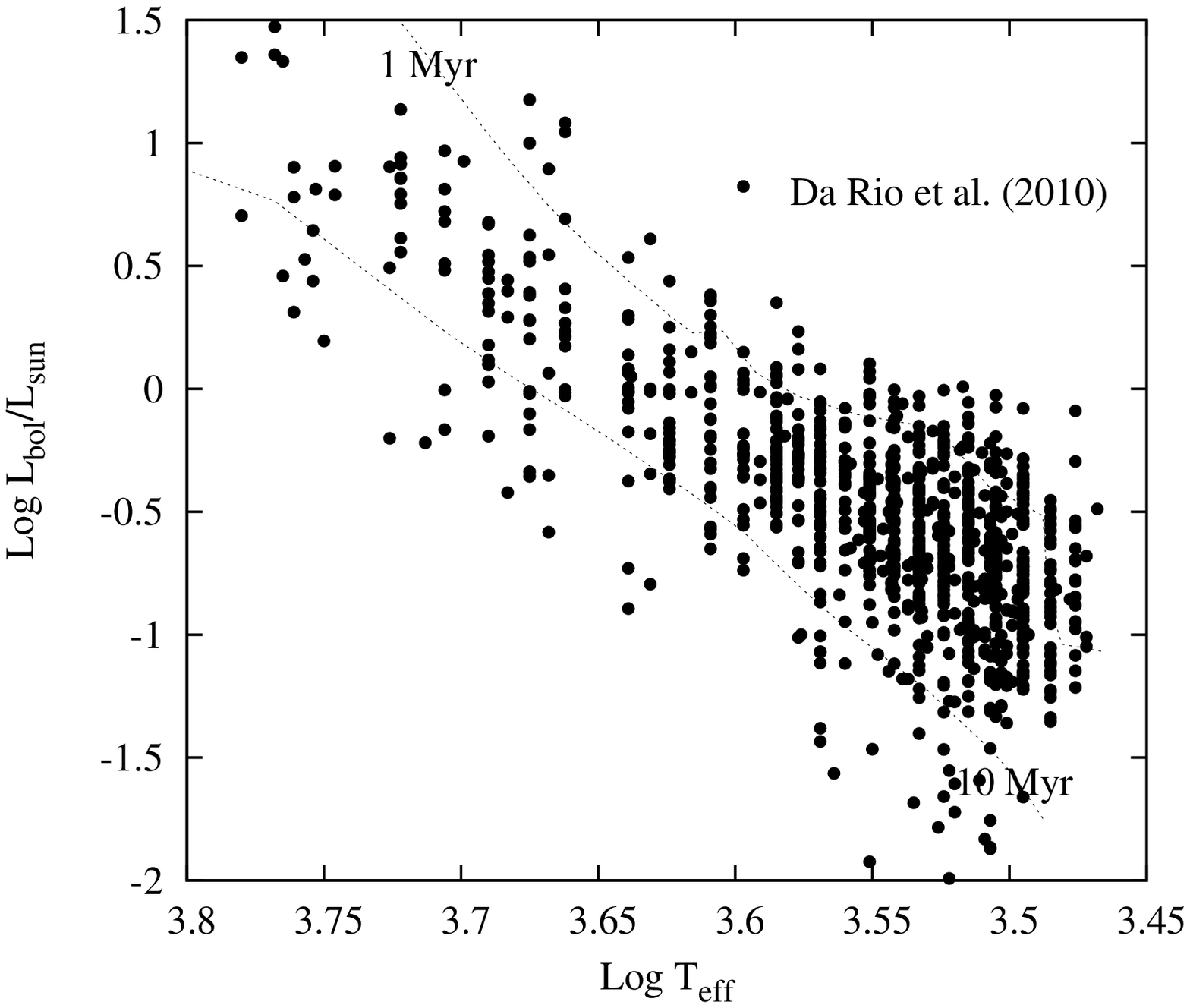}
\includegraphics[width=80mm]{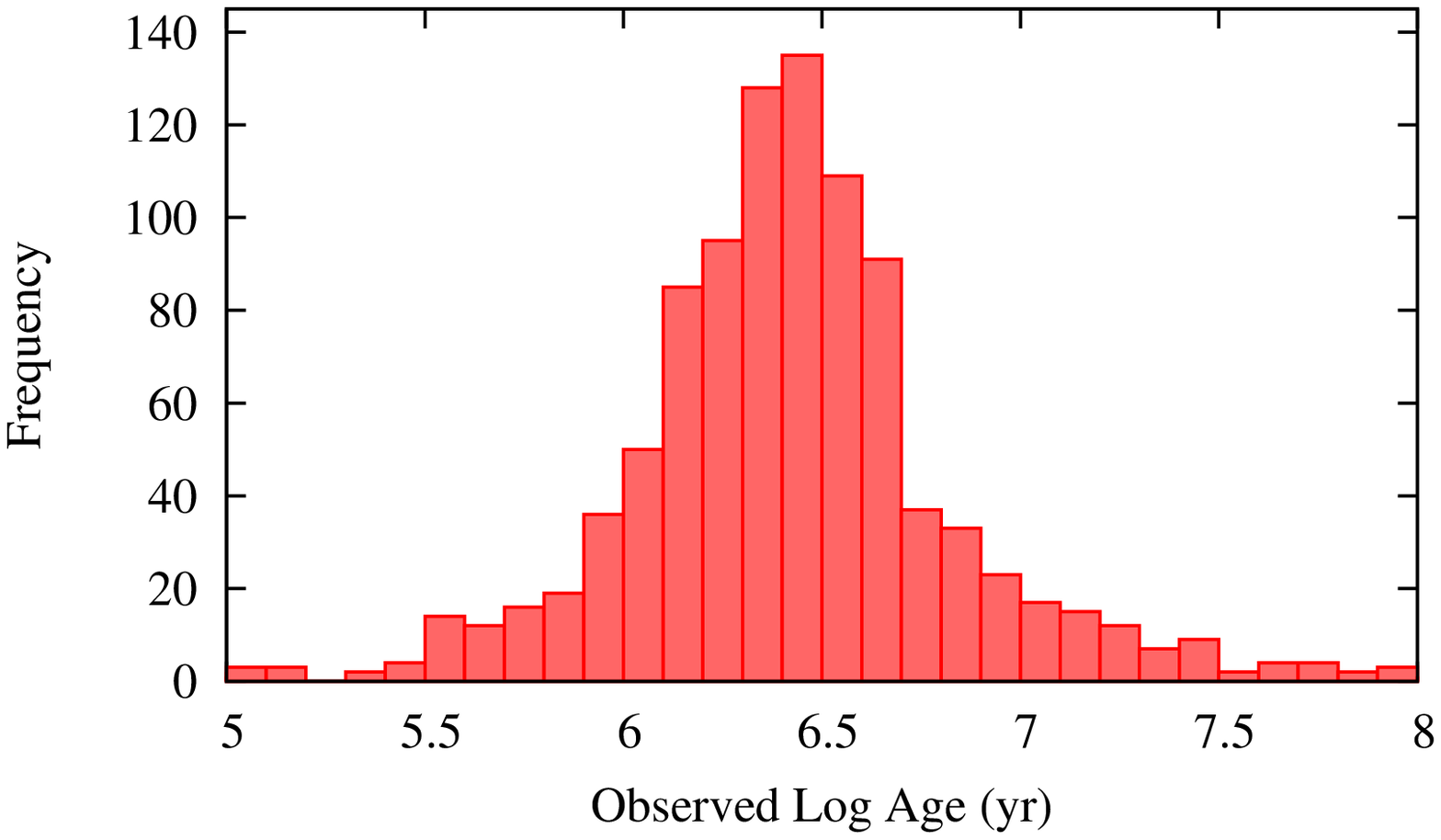}

\caption{(Top) The Hertzsprung-Russell (HR) diagram for stars in the Orion
  Nebula cluster included in the catalogue compiled by Da Rio et
  al. (2010b). The dashed loci are 1 and 10\,Myr isochrones from Siess
  et al. (2000).  (Bottom) The inferred distribution of $\log$ age (in
  years).}
\label{onchr}
\end{figure}

The goal of this investigation is to search for differences in the age
distributions of stars with and without discs in a large and
homogeneous sample from the ONC, and thus estimate the extent of any
real age spread. The observational basis is the optical catalogue and
HR diagram of the ONC produced by DR10. This improves on earlier work
by \citet{hillenbrand97} and is the largest homogeneous catalogue of
photometry and spectral types for stars in the ONC. DR10 simultaneously
used photometry and spectroscopy to estimate extinction and accretion
luminosity and hence find the intrinsic bolometric luminosity and
effective temperature for each source. The sample of ONC stars was
filtered to exclude possible non-members with membership probabilities
based on proper motion that were smaller than 50\%.\footnote{As
discussed in DR10, there are very few objects with membership
probabilities between 10\% and 90\%, so the exact choice of membership
criterion is not important} The overall level of non-member
contamination remaining in the catalogue was estimated to be 2--3\%.

DR10 used the evolutionary models of \citet{palla99} and
\citet{siess00} to assign masses and ages from the resultant HR diagram
and to study their distributions.  They preferred the models of
\citet{siess00} as they produced a smaller apparent age spread and a
mean age that was independent of stellar mass. The catalogue is
incomplete due to (a) lack of photometry or spectral types, which
becomes severe at $V>18$ and (b) crowding in the inner parts of the
ONC. This affects stars at all magnitudes, but is more serious for
fainter objects. For similar extinctions this of course means that
incompleteness becomes greater for less luminous objects (see Fig.~19
of DR10) and, for the purposes of this investigation, biases the sample
against ``older'' stars (where ``older'' means as judged from position
in the HR diagram).

The 976 stars from the catalogue of DR10, that have ages deduced from
the model isochrones, are plotted in a HR diagram in
Fig.~\ref{onchr}a. The inferred age distribution is shown as a
histogram in Fig.~\ref{onchr}b. The overall age distribution can be
approximated as log-normal, with a mean of 6.42 and a dispersion
(1-sigma) of 0.43 dex, although there is some kurtosis and it could be
argued that some sort of core plus halo distribution is more
appropriate (see Section 4.1).  The mean and median masses of this
sample are $0.50\,M_{\odot}$ and $0.32\,M_{\odot}$ respectively, and
there is no significant correlation between mass and age.

The catalogue of DR10 was cross-correlated against three independent
catalogues of ONC data that enable a judgement as to whether the stars
possess a circumstellar disc. These catalogues are:
\begin{enumerate}
\item Stars in the ONC with measurements of near-infrared excess in the
  $I-K$ colour \citep{hillenbrand98}. 
  The cross-matched catalogue contains 535 stars spread over
  about 700 square arcminutes, with some concentration towards the
  central Trapezium region. The same criterion used by
  \citet{hillenbrand98}
 was adopted to signal the presence of a disc; namely that $I-K$ is more than
  0.3 mag redder than expected from the photospheric spectral type. Using this
  criterion there are 295 stars with discs and 240 without. This sample
  is subsequently referred to as the ``$K$-band sample''.

\item Stars in the central regions of the ONC with a measurement of
  near infra-red excess using $JHKL$ data \citep{lada00}. 
  The cross-matched catalogue in this
  case consists of 150 objects in a much smaller 36 square arcminute
  region surrounding the Trapezium. Discs are identified in the $J-H$
  versus $K-L$ diagram using the same criterion adopted by \citet{lada00};
  namely that the star's $K-L$ colour lies redward of a reddening line
  extending from the intrinsic colours of a low-mass main-sequence star
  \citep[defined by][]{bessell88}. Using this criterion there are
  115 stars with discs and 35 without. This sample is
  subsequently referred to as the ``$L$-band sample''.

\item Stars in the ONC with measurement of infrared excess using the
  Spitzer [3.6]-[8.0] colour (Megeath et al. in prep.). The
  cross-matched catalogue contains 425 objects spread over about 1300
  square arcminutes, with no obvious concentration in the central
  Trapezium region. The relatively simple criterion of
  whether the [3.6]-[8.0] colour is greater than 0.7 mag is adopted to
  diagnose the presence of a disc \citep[see][]{cieza06}. Using
  this criterion there are 290 stars with discs and 135 without. This
  sample is subsequently referred to as the ``Spitzer sample''.

\end{enumerate}
Each of these samples is subject to a variety of selection effects,
which are discussed further subsequently, but all three have a high
fraction of stars less massive than the Sun (86\%, 77\% and 84\% respectively).

\section{Comparison of stars with and without discs}

\begin{figure*}
\centering
\begin{minipage}[t]{0.33\textwidth}
\includegraphics[width=60mm]{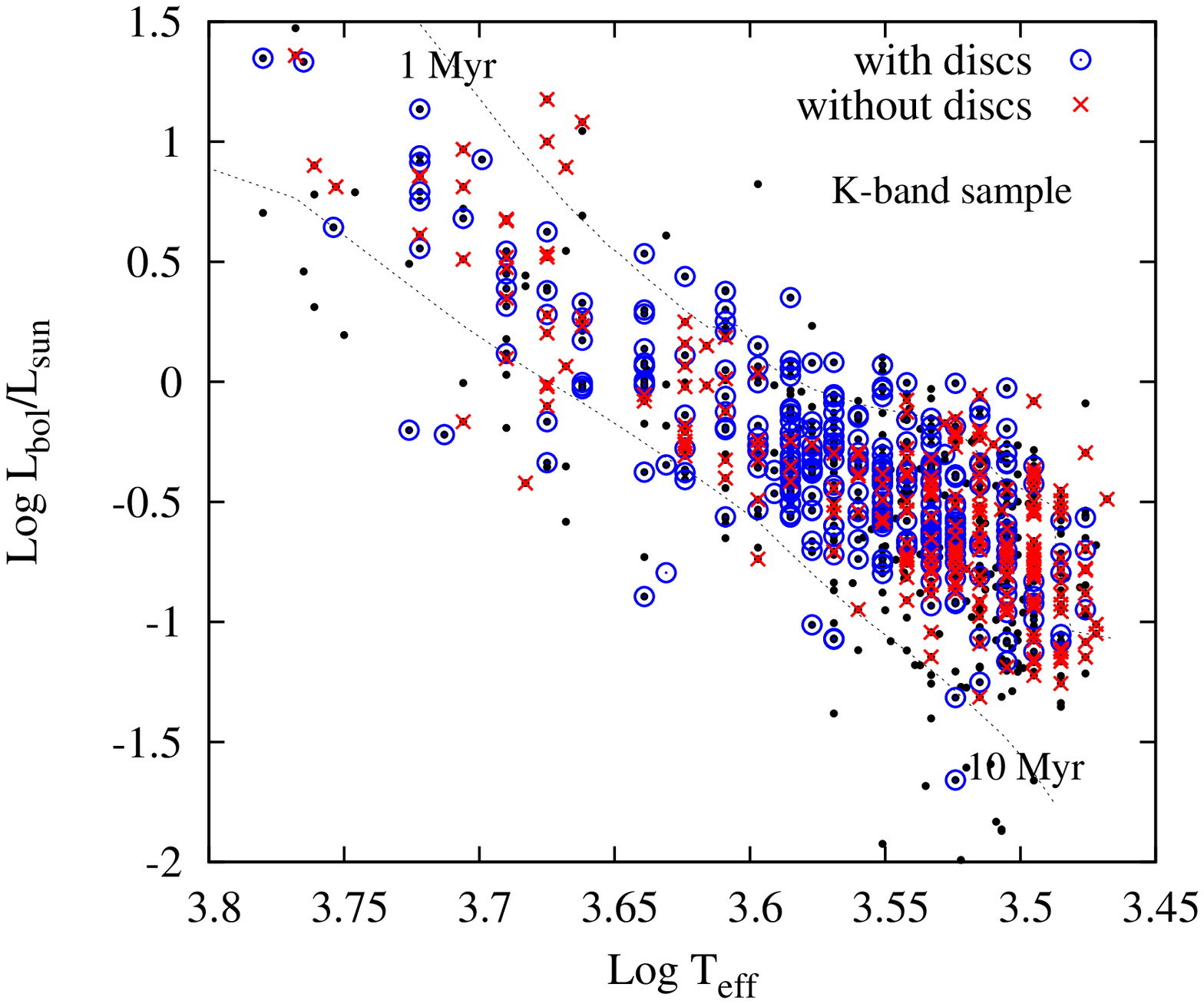}
\includegraphics[width=60mm]{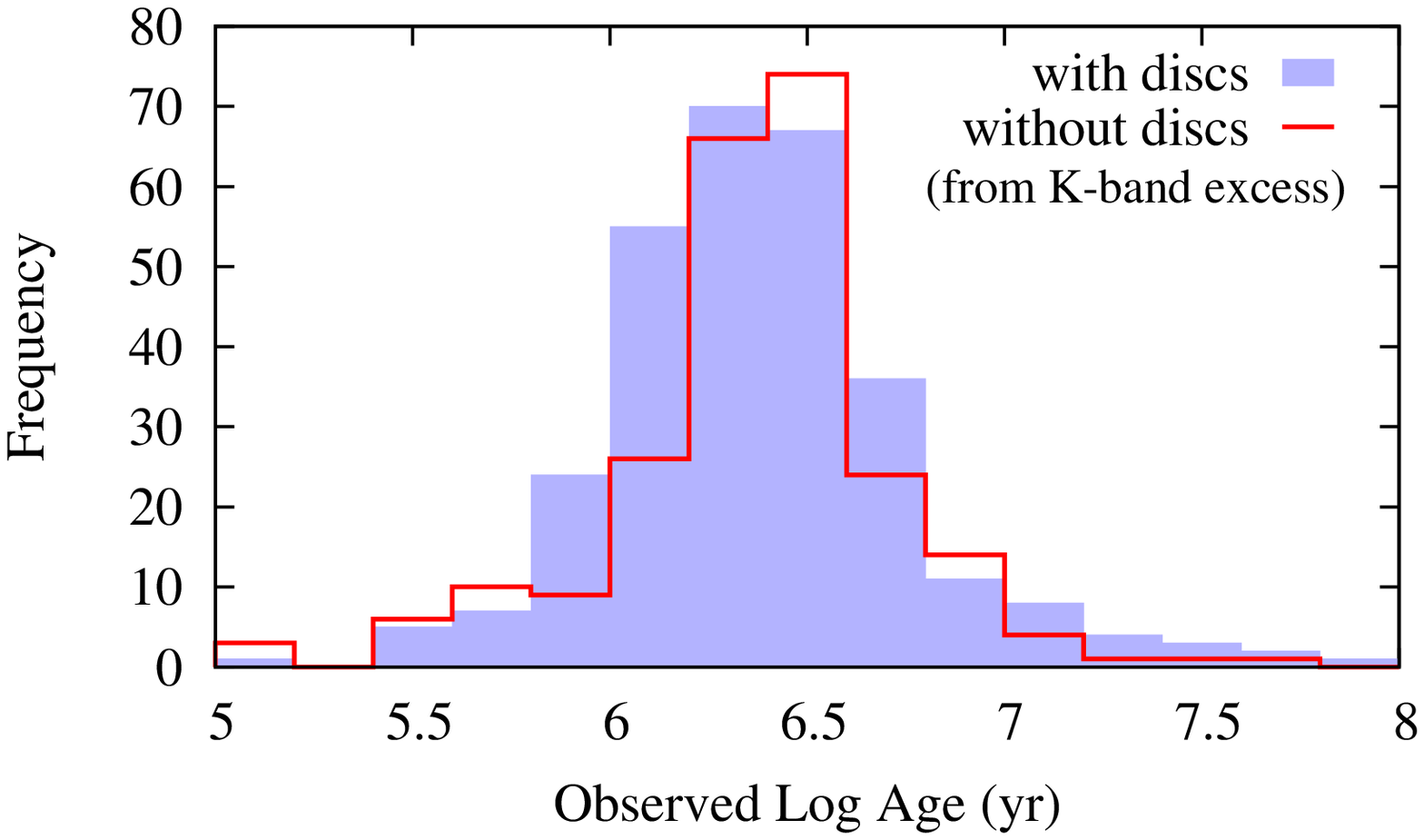}
\end{minipage}
\begin{minipage}[t]{0.33\textwidth}
\includegraphics[width=60mm]{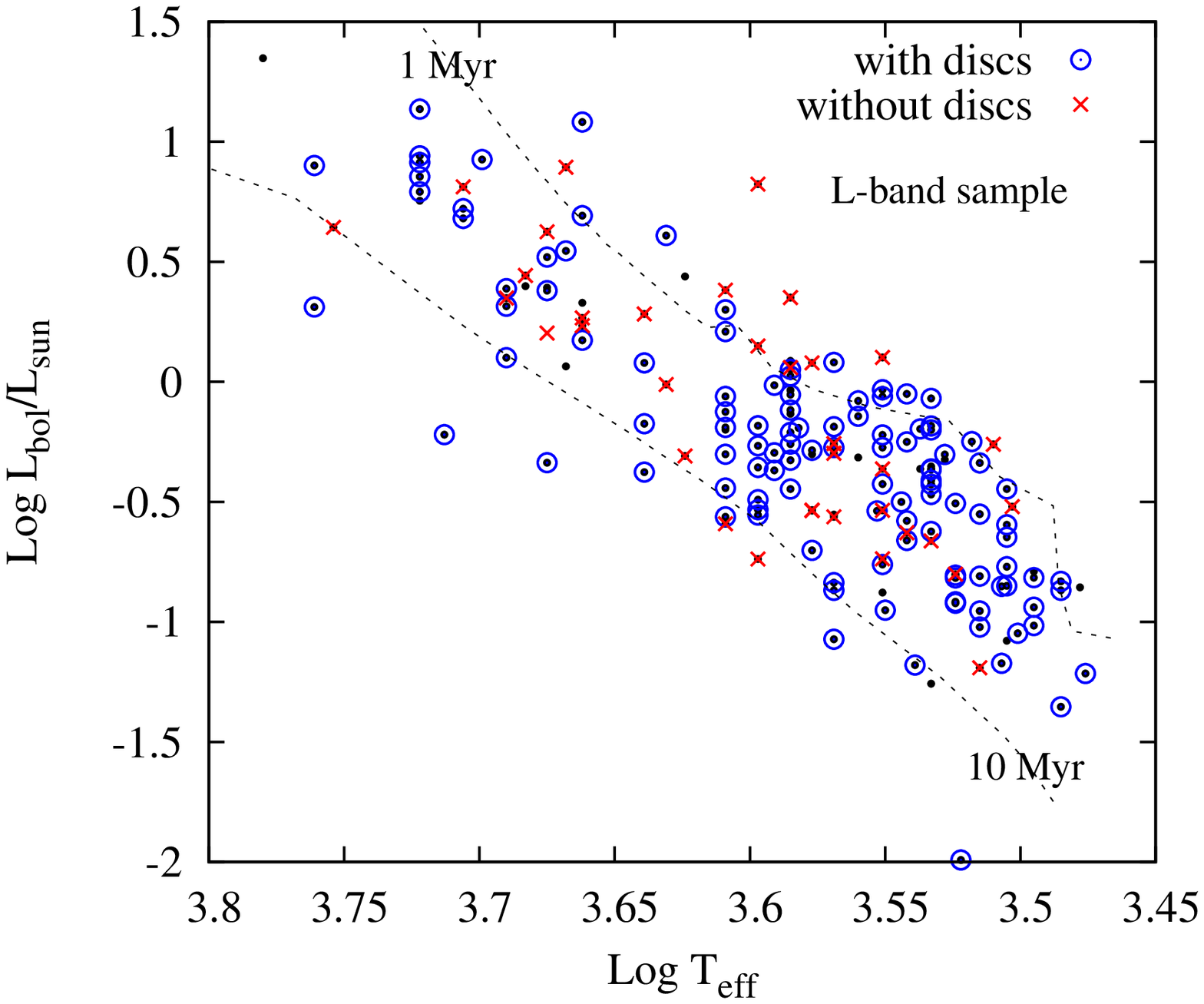}
\includegraphics[width=60mm]{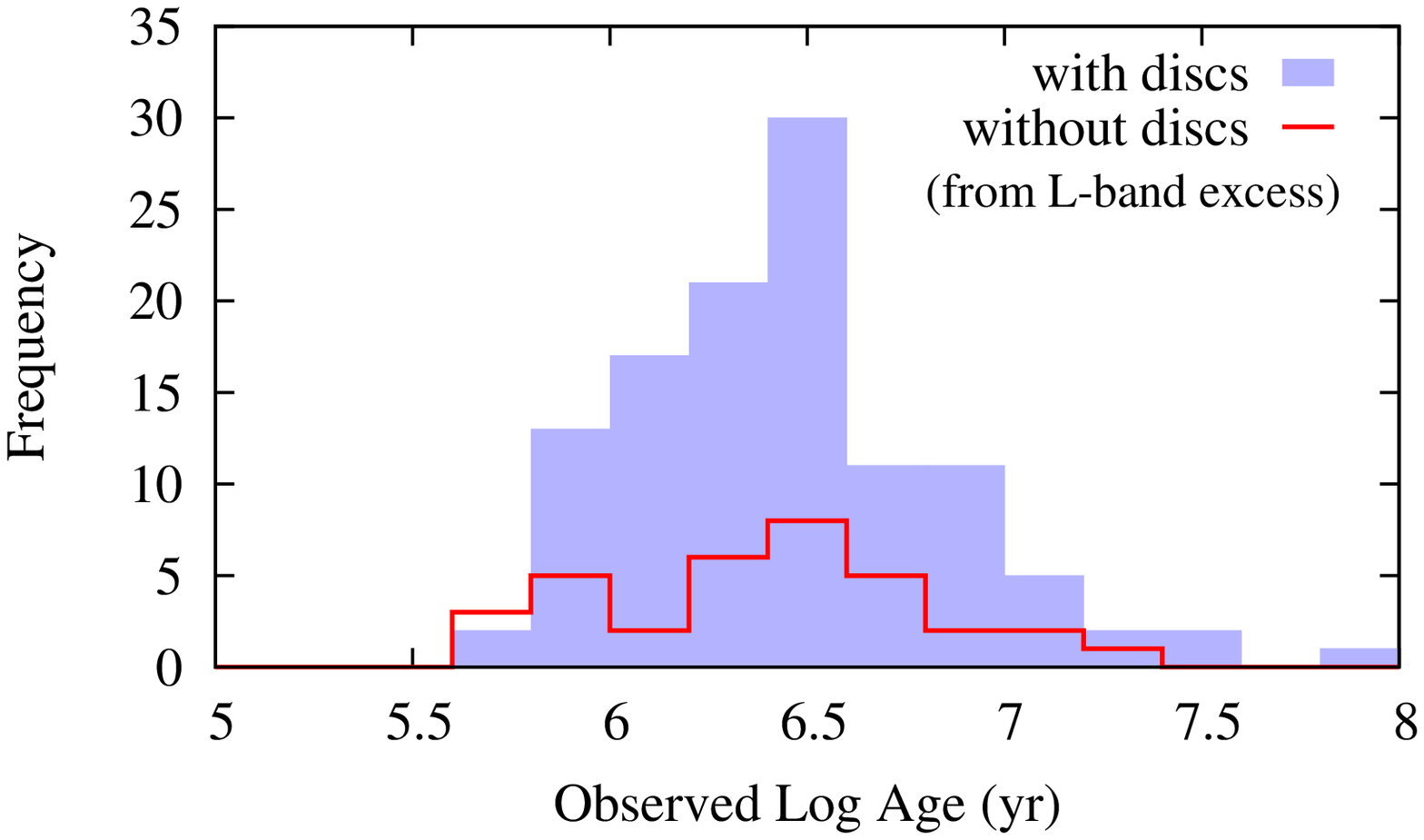}
\end{minipage}
\begin{minipage}[t]{0.33\textwidth}
\includegraphics[width=60mm]{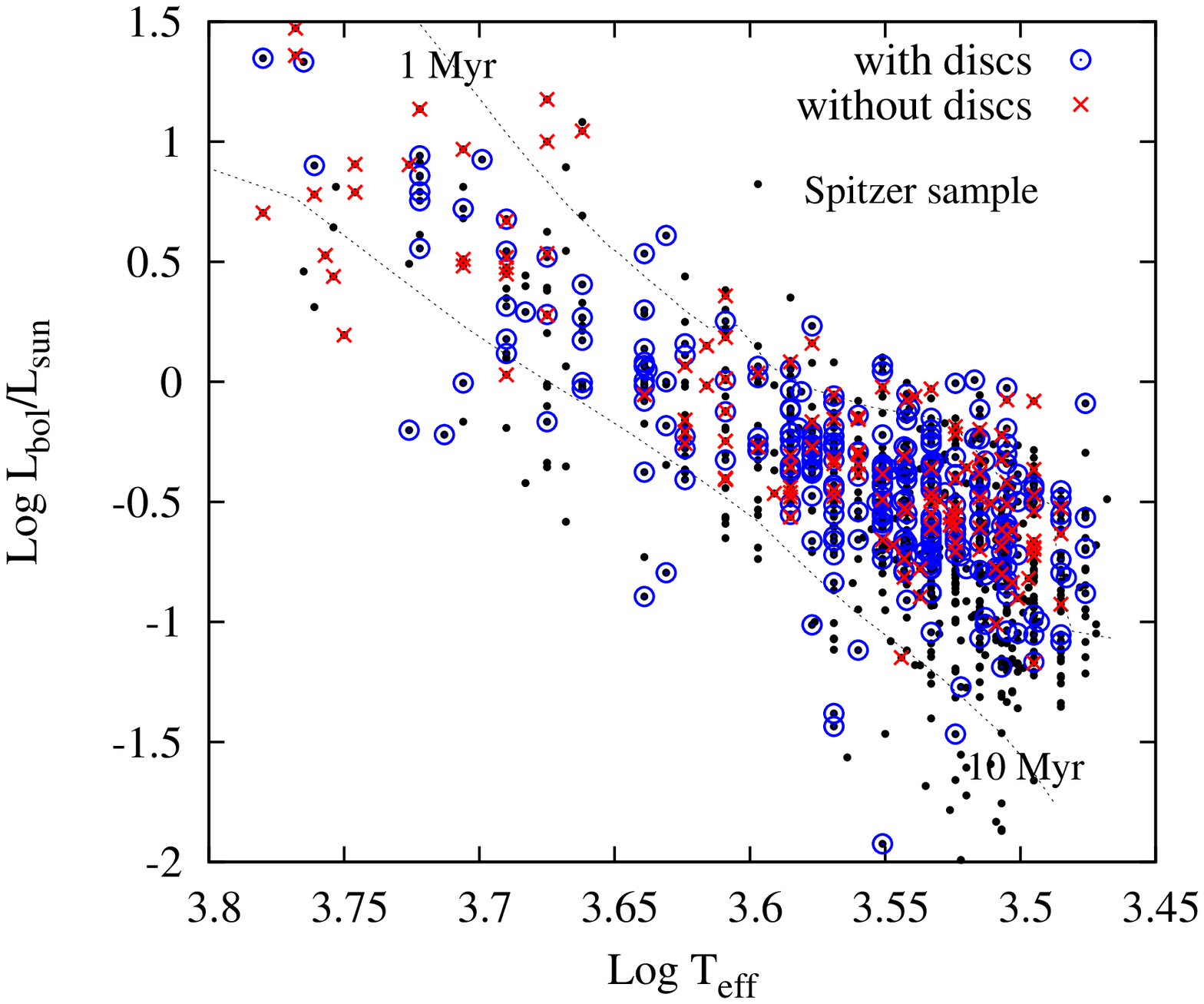}
\includegraphics[width=60mm]{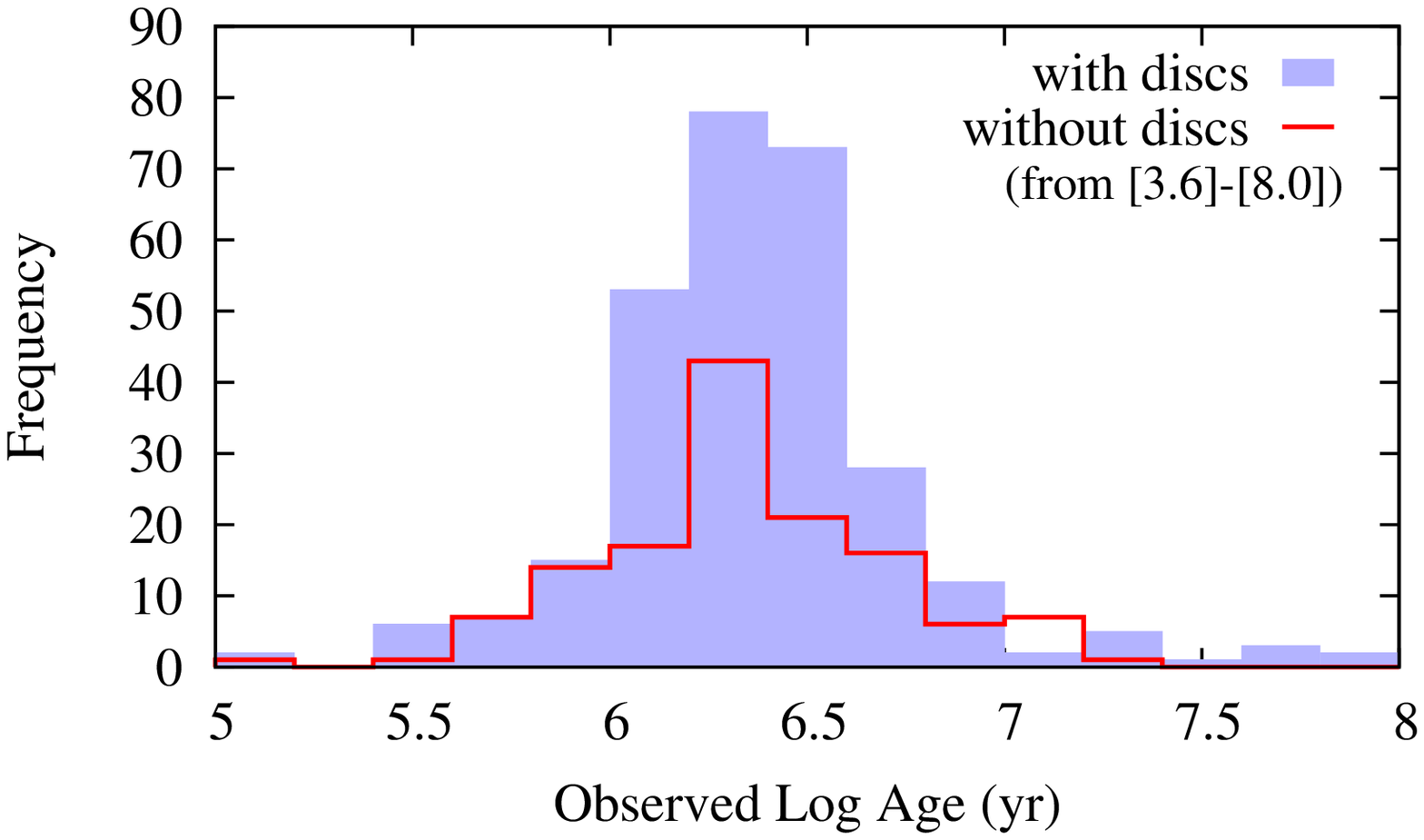}
\end{minipage}

\caption{(Top row): Hertzsprung-Russell (HR) diagrams showing the locations
  of stars with and without discs, diagnosed using three different
  methods. Isochrones at 1 and 10\,Myr are shown from the models of
  Siess et al. (2000). To demonstrate the level of completeness, 
  the parent sample stars from Da Rio et al. (2010b) that
  lie within the area covered by each of the infrared surveys is
  shown with small dots. 
(Bottom Row:) Histograms of the age
  distributions inferred from the HR diagrams for stars with and without discs.
}
\label{hrplot}
\end{figure*}

\begin{table*}

\caption{A comparison of the bulk properties of the parent sample of
  ONC stars from the Da Rio et al. (2010b) catalogue with subsamples
  that have the presence of discs diagnosed by three different methods.
  For each of the three subsamples the table also shows the 
  probability (judged by a two-tailed Kolmogorov-Smirnov test) 
  that the null hypotheses, that the age and mass
  distributions of stars with and without discs are drawn from the same
  parent distribution, can be rejected.}

\begin{tabular}{lccccccccc}
\hline
 Sample & Size  && \multicolumn{3}{c}{$\log$ Age (years)} &
 & \multicolumn{3}{c}{Mass ($M_{\odot}$)} \\
        &       &\hspace*{5mm}&  Mean & Std. Dev. & Median & \hspace*{5mm}& Mean & Std. Dev. &
 Median \\
\hline
Dario et al. (2010b) & 976 && 6.42 & 0.43 & 6.41&& 0.50 & 0.47 & 0.32 \\
parent sample       &     &&      &      &     &&      &      &      \\

&&&&&&&&&\\
$K$-band sample, Hillenbrand et al. (1998) &&&&&&&&&\\
 All with $I-K$     & 535 && 6.37 & 0.38 & 6.38 && 0.57 & 0.53 & 0.36 \\
 With Discs         & 295 && 6.37 & 0.38 & 6.37 && 0.59 & 0.48 & 0.42 \\
 Without Discs      & 240 && 6.36 & 0.37 & 6.40 && 0.55 & 0.58 & 0.30 \\
&&&&&&&&&\\
Null KS Probability      &&     & \multicolumn{3}{c}{0.90} &&
 \multicolumn{3}{c}{$>0.9999$} \\
With vs Without\\
&&&&&&&&&\\
$L$-band sample, Lada et al. (2000) &&&&&&&&&\\
All with $JHKL$     & 150 && 6.42 & 0.42 & 6.41 && 0.80 & 0.63 & 0.56 \\
With Discs          & 115 && 6.45 & 0.40 & 6.43 && 0.72 & 0.62 & 0.46 \\
Without Discs       &  35 && 6.34 & 0.47 & 6.41 && 0.84 & 0.57 & 0.56 \\
&&&&&&&&&\\
Null KS Probability      &&     & \multicolumn{3}{c}{0.73} &&
 \multicolumn{3}{c}{0.90} \\
With vs Without\\
&&&&&&&&&\\
Spitzer sample, Megeath et al. (in prep.) &&&&&&&&\\
All with [3.6]-[8.0] & 425 && 6.35 & 0.41 & 6.34 && 0.60 & 0.56 & 0.37 \\
With Discs           & 290 && 6.36 & 0.42 & 6.36 && 0.55 & 0.50 & 0.36 \\
Without Discs        & 135 && 6.33 & 0.39 & 6.33 && 0.70 & 0.66 & 0.39 \\
&&&&&&&&&\\
Null KS Probability      &&     & \multicolumn{3}{c}{0.51} &&
 \multicolumn{3}{c}{0.96} \\
With vs Without\\
\hline
\end{tabular}
\label{tabhrresults}
\end{table*}

Figure~\ref{hrplot} shows the HR diagrams and inferred ``age
distributions'' using the three sources of information on disc
presence. In each case, the stars with and without discs are identified
and the parent sample from the catalogue of DR10 that
share the same spatial extent as each survey is shown to illustrate
completeness. Table~\ref{tabhrresults} reports the means and
standard deviations of the distributions of $\log$ age for each
subsample as well as the median mass derived by DR10.

The $K$-band and Spitzer samples contain stars down to lower masses
than the $L$-band sample because they are more sensitive in absolute
terms in detecting stellar photospheres. Paradoxically,
Fig.~\ref{hrplot} shows that the $L$-band sample is almost complete in
terms of providing data for the parent sample of stars from the DR10
catalogue, whereas there is some incompleteness at fainter magnitudes
and lower luminosities in the $K$-band and Spitzer samples. The reason
for this is that the DR10 sample is also more incomplete for faint
stars near the centre of the ONC where the $L$-band sample is, but
achieves greater sensitivity in the outer regions where most of the
$K$-band and Spitzer sample stars are.

It is clear from Fig.~\ref{hrplot} and Table~\ref{tabhrresults} that
the age distributions of stars with and without discs are similar using
each of the three methods of identifying discs. The means and variances
of the disc/no-disc subsamples are judged to be not significantly
different in each case, using T-tests and F-tests respectively
\citep[see][]{press92}, although
there are small differences in the overall mean age between the three
samples.

Whether the age distributions of stars with and without discs are drawn
from the same parent distribution, was tested with two-sided
Kolmogorov-Smirnov (K-S) tests on the cumulative age distributions
\citep{press92}. Similar tests were performed on the cumulative mass
distributions. The results of these are reported in
Table~\ref{tabhrresults} in terms of a probability that the null
hypothesis (that the two distributions are drawn from the same
population) can be rejected. The age distributions are
indistinguishable in the cases of the $L$-band and Spitzer samples and
are only marginally different at a 90 per cent confidence level in the
case of the $K$-band sample.  The mass distributions are not
significantly different for stars with and without discs diagnosed in
the $L$-band, are marginally different for the Spitzer samples (the
median masses are similar but there is an excess of tail of higher mass
stars in the discless population), but there is a highly significant
difference in the mass distributions of stars with and without discs in
the $K$-band sample -- the stars with discs have a higher median
mass. It is worth noting that the K-S tests we use are very insensitive
to outliers in the distributions (see Press et al. 1992).

The $K$-band disc census is likely to be quite
incomplete. It is well known \citep[see][]{hillenbrand98, lada00}
that the effectiveness of $K$-band excess measurements
reduces with decreasing mass because the contrast between the
photosphere and warm dust diagnosed in the $K$-band becomes
smaller. This probably accounts for the lower disc frequency in the
$K$-band sample compared to the $L$-band and Spitzer samples,
 and likely accounts for the differing mass
distributions of the stars with and without discs. Such incompleteness
could also bias the comparison of the age distributions, because the disc
lifetime for lower mass stars may be longer 
\citep{carpenter06, kennedy09}.
In this case the average age of stars with discs would be biased
downwards by being unable to detect longer-lived discs in the lower
mass stars, whilst the average age of stars without discs (or at least
appearing to have no disc) would
consequently be biased upwards.  This selection effect is likely to
be much weaker
for discs detected by excesses in the $L$-band or at 8\,$\mu$m and
in any case the mass distributions of stars with and without discs in these
samples are not very different.

Irrespective of these complications we have found no significant
evidence (even using $K$-band excesses) that stars with discs are
younger than stars without discs, which is consistent with the highly
overlapping locations of of these samples in the HR diagrams (see
Fig.~\ref{hrplot}).  This appears to contradict the idea that most
stars are born with discs and then lose them on timescales that are
comparable to or shorter than the claimed spread of ages in the ONC and
is instead consistent with the idea that the stellar population has an age
spread smaller than a typical disc lifetime. However, a quantitative
treatment needs to consider the influence of observational
uncertainties and any other physical mechanisms that might cause
scatter in the HR diagram.

\section{An interpretive model}

\begin{figure*}
\centering
\begin{minipage}[t]{0.33\textwidth}
\includegraphics[width=55mm]{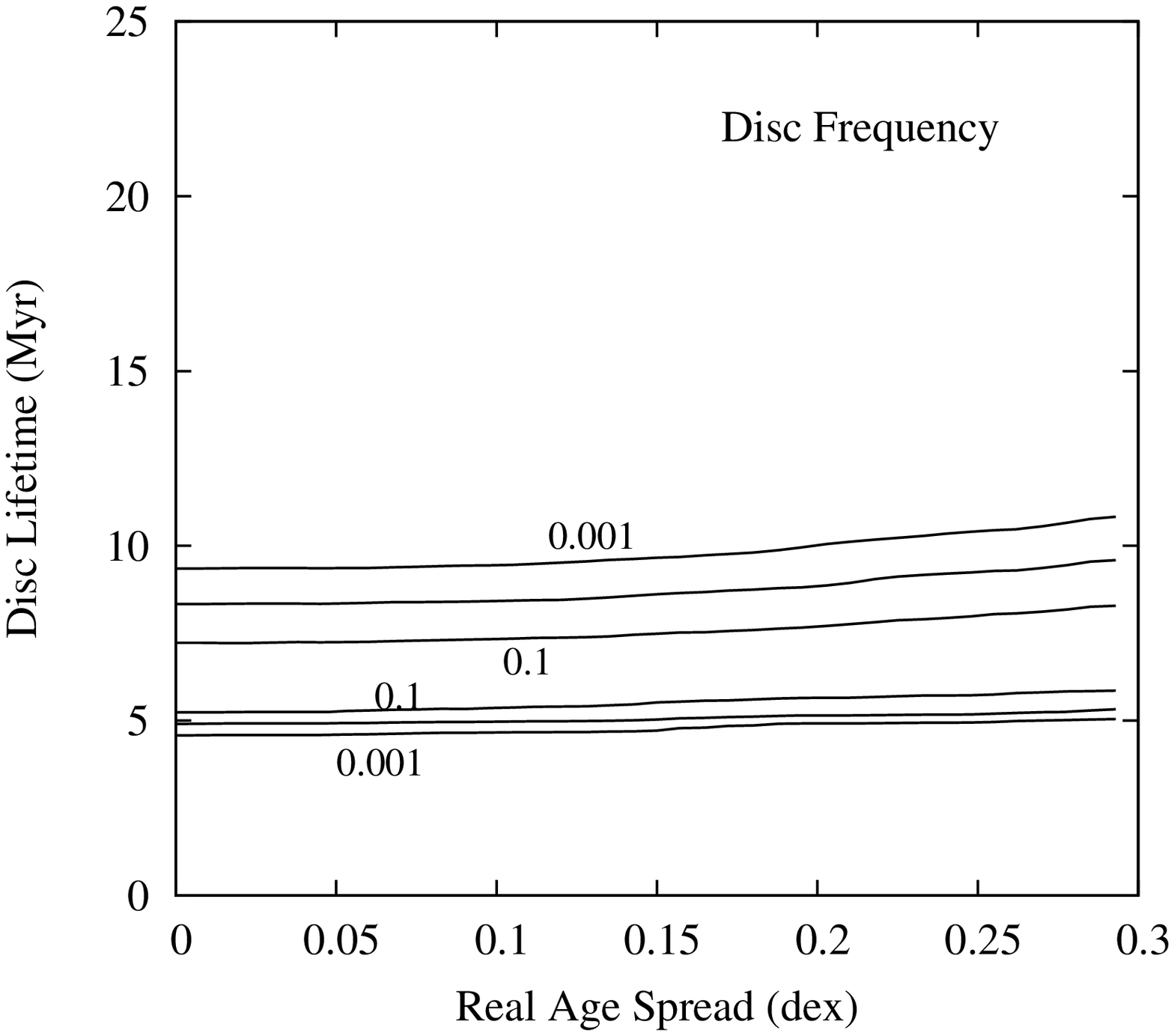}
\end{minipage}
\begin{minipage}[t]{0.33\textwidth}
\includegraphics[width=55mm]{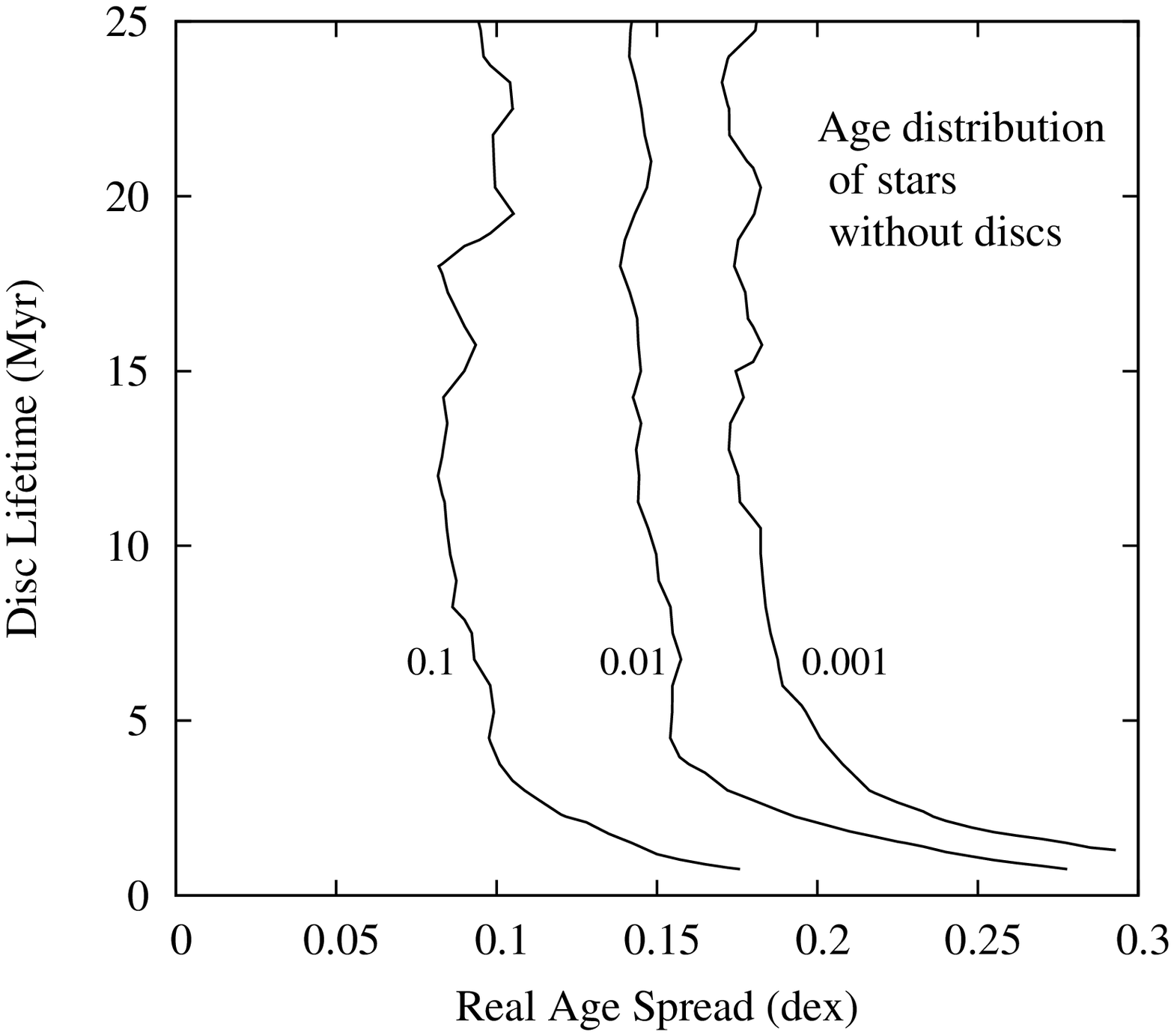}
\end{minipage}
\begin{minipage}[t]{0.33\textwidth}
\includegraphics[width=55mm]{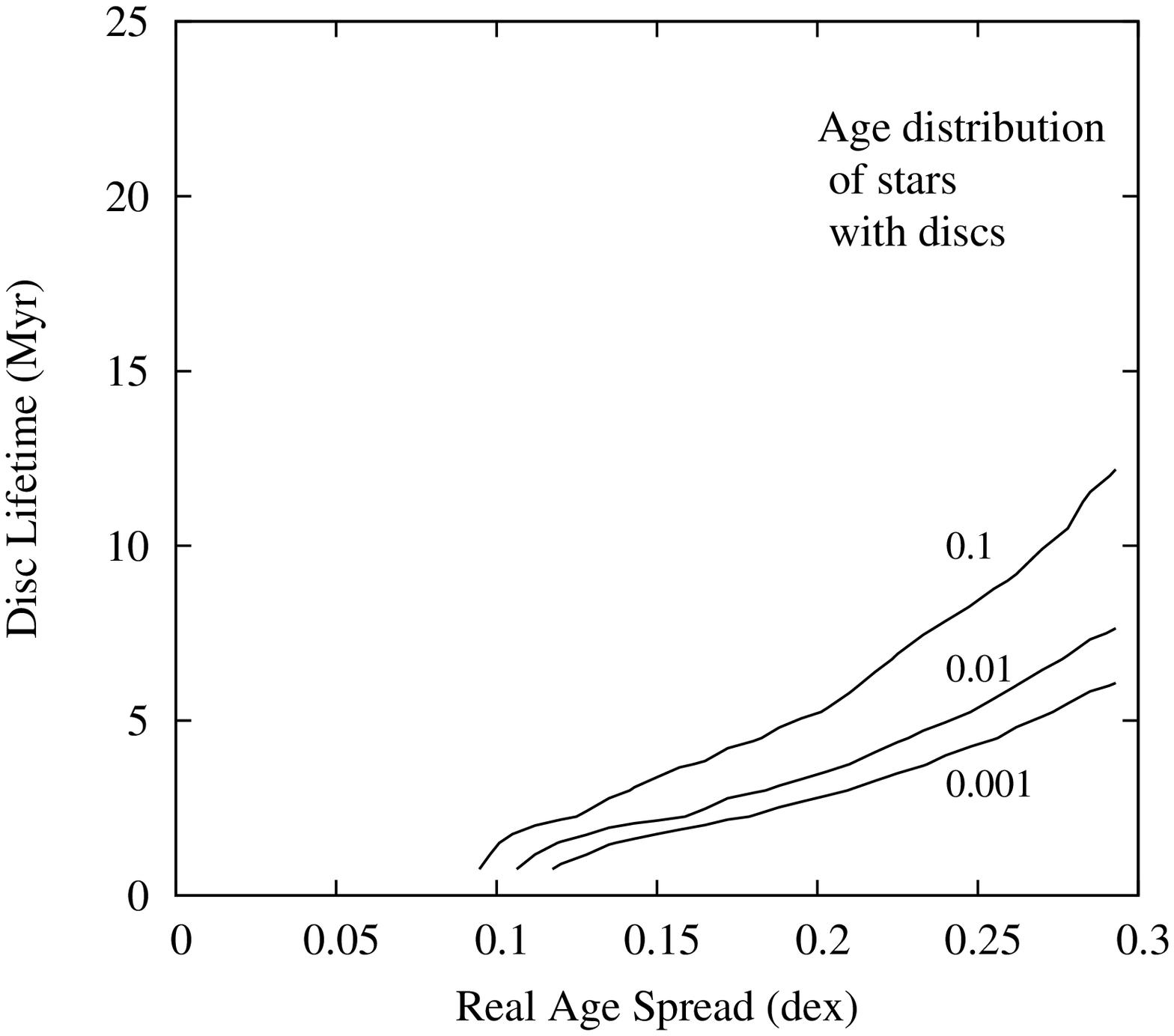}
\end{minipage}

\caption{Contour plots of probability that the models described in
 Section~4.1 provide a good fit to the properties of the Spitzer
 sample (plots for the other two samples are qualitatively similar): 
the disc frequency (left); the apparent age distribution of
 stars without discs (centre); and the apparent age distribution of
 stars with discs (right).  In each plot the contours represent
 probability levels of 0.1, 0.01 and 0.001 respectively.
 }
\label{probgrids}
\end{figure*}

\begin{figure*}
\centering
\begin{minipage}[t]{0.33\textwidth}
\includegraphics[width=60mm]{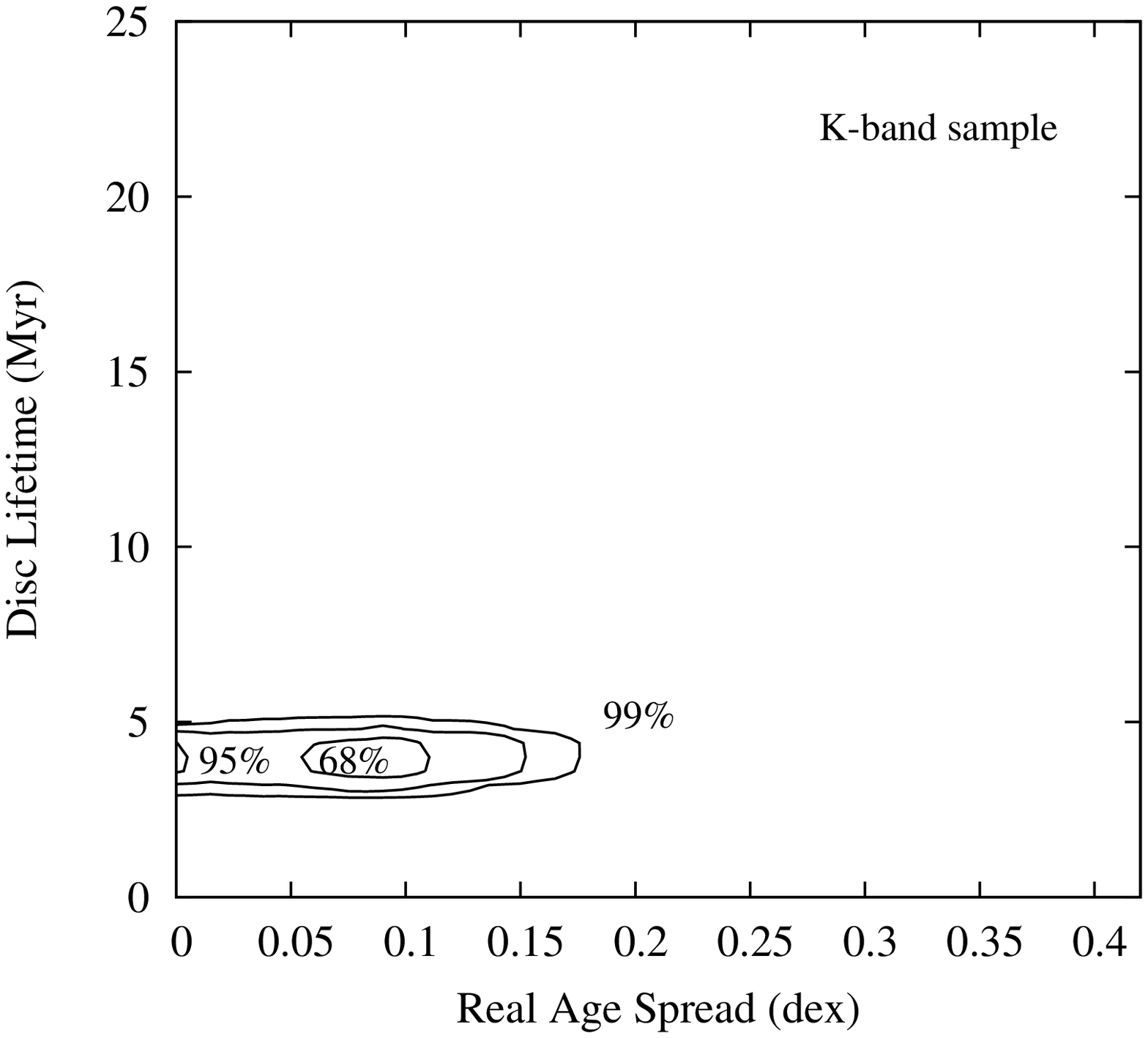}
\includegraphics[width=60mm]{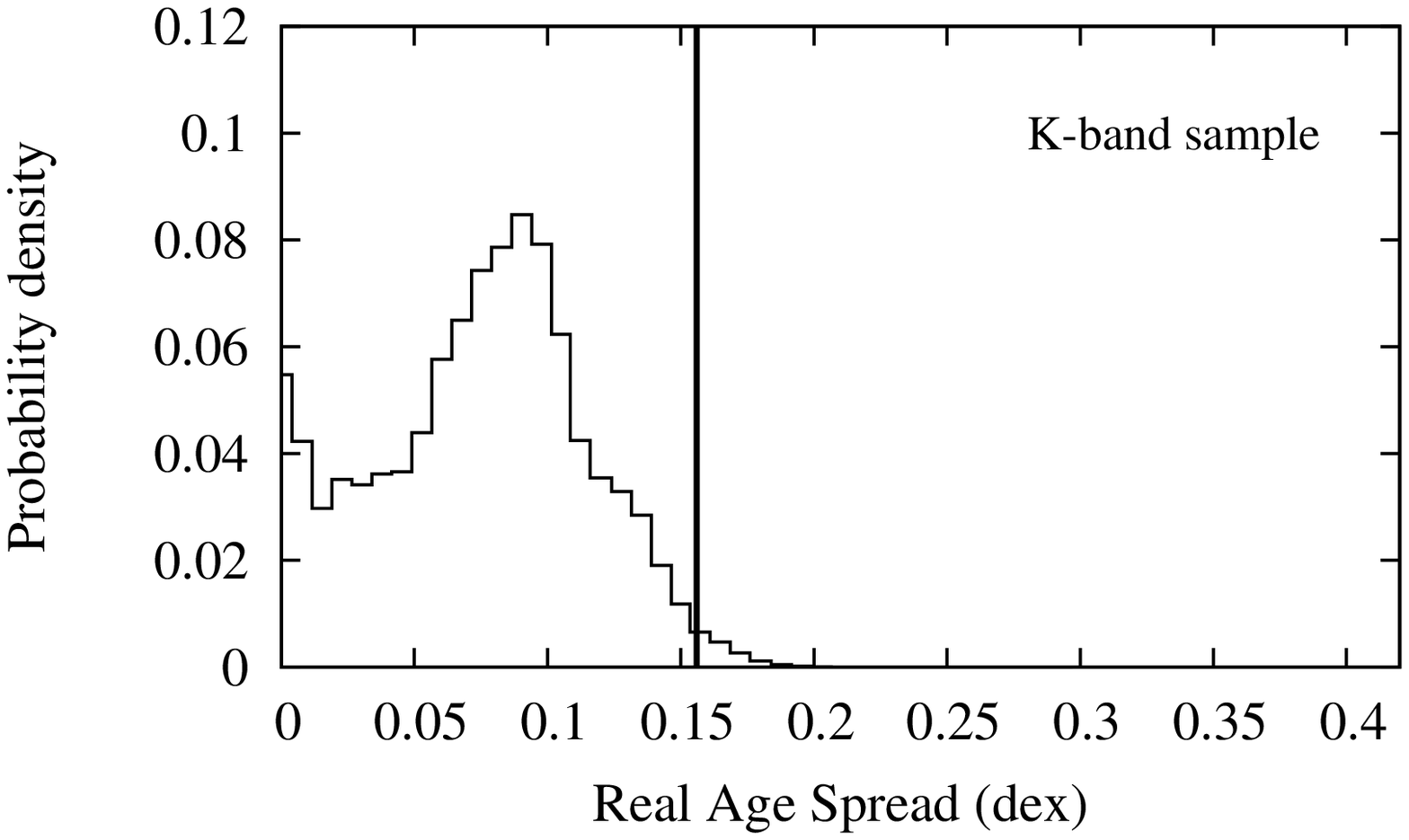}
\includegraphics[width=60mm]{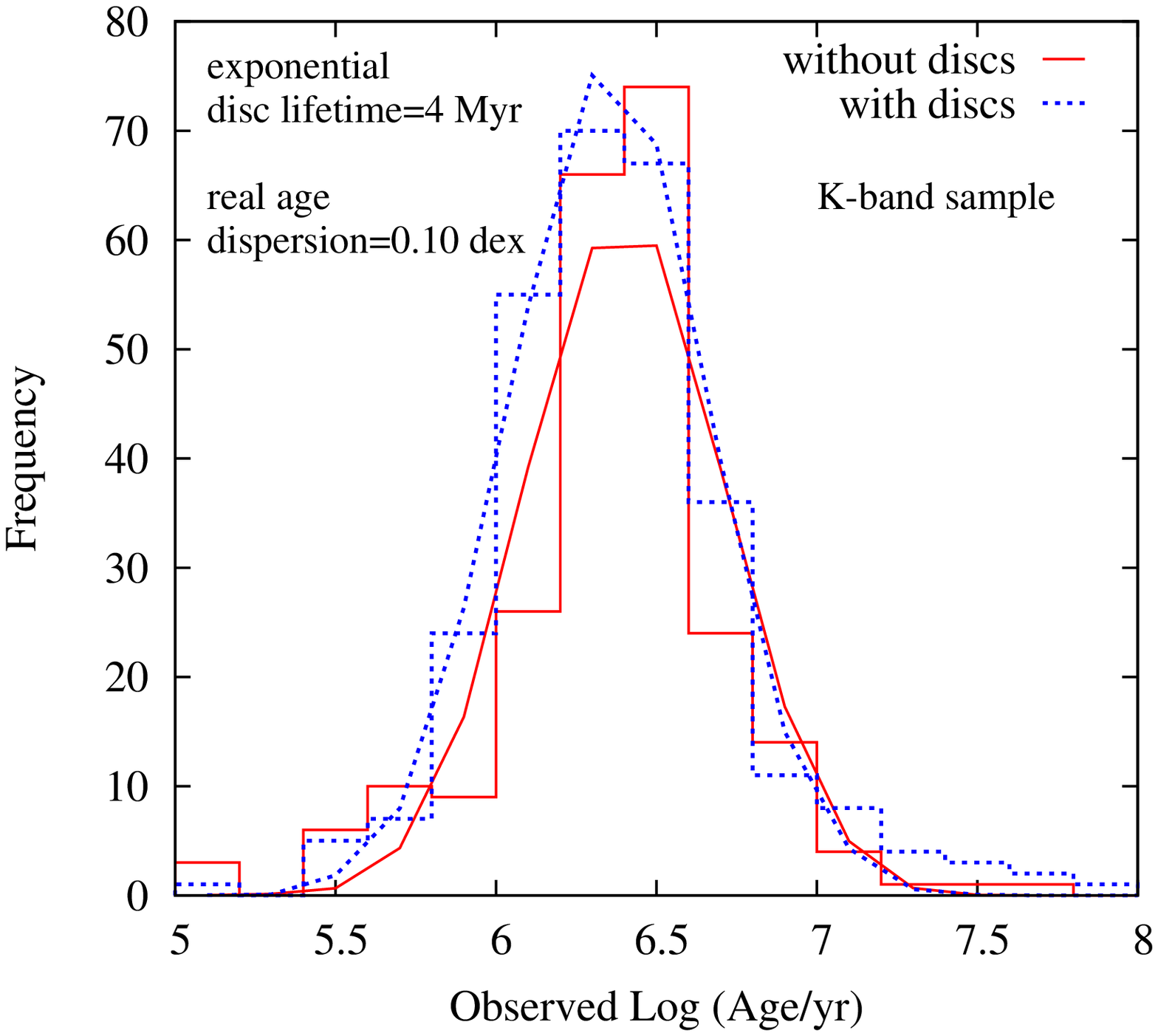}
\end{minipage}
\begin{minipage}[t]{0.33\textwidth}
\includegraphics[width=60mm]{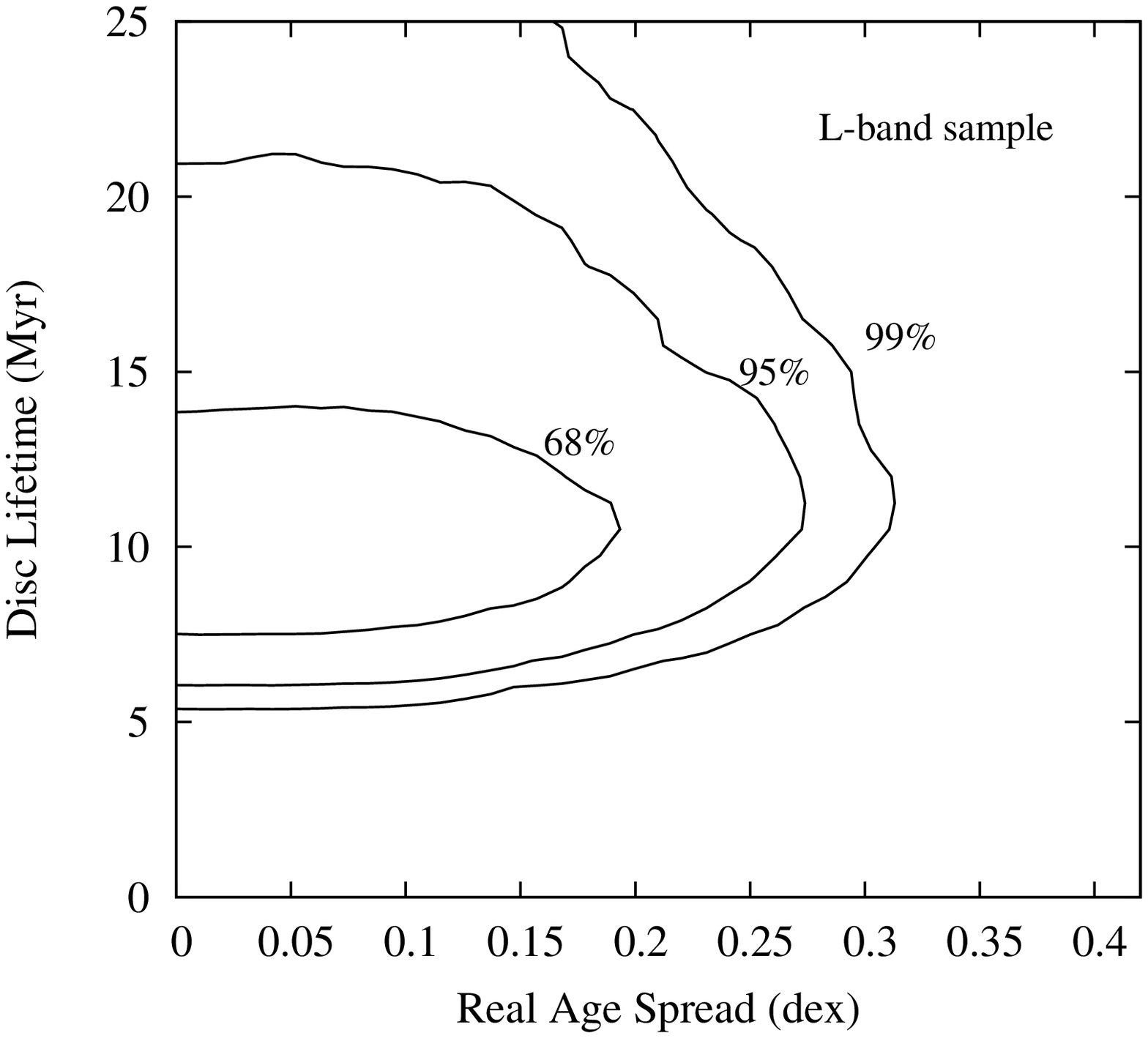}
\includegraphics[width=60mm]{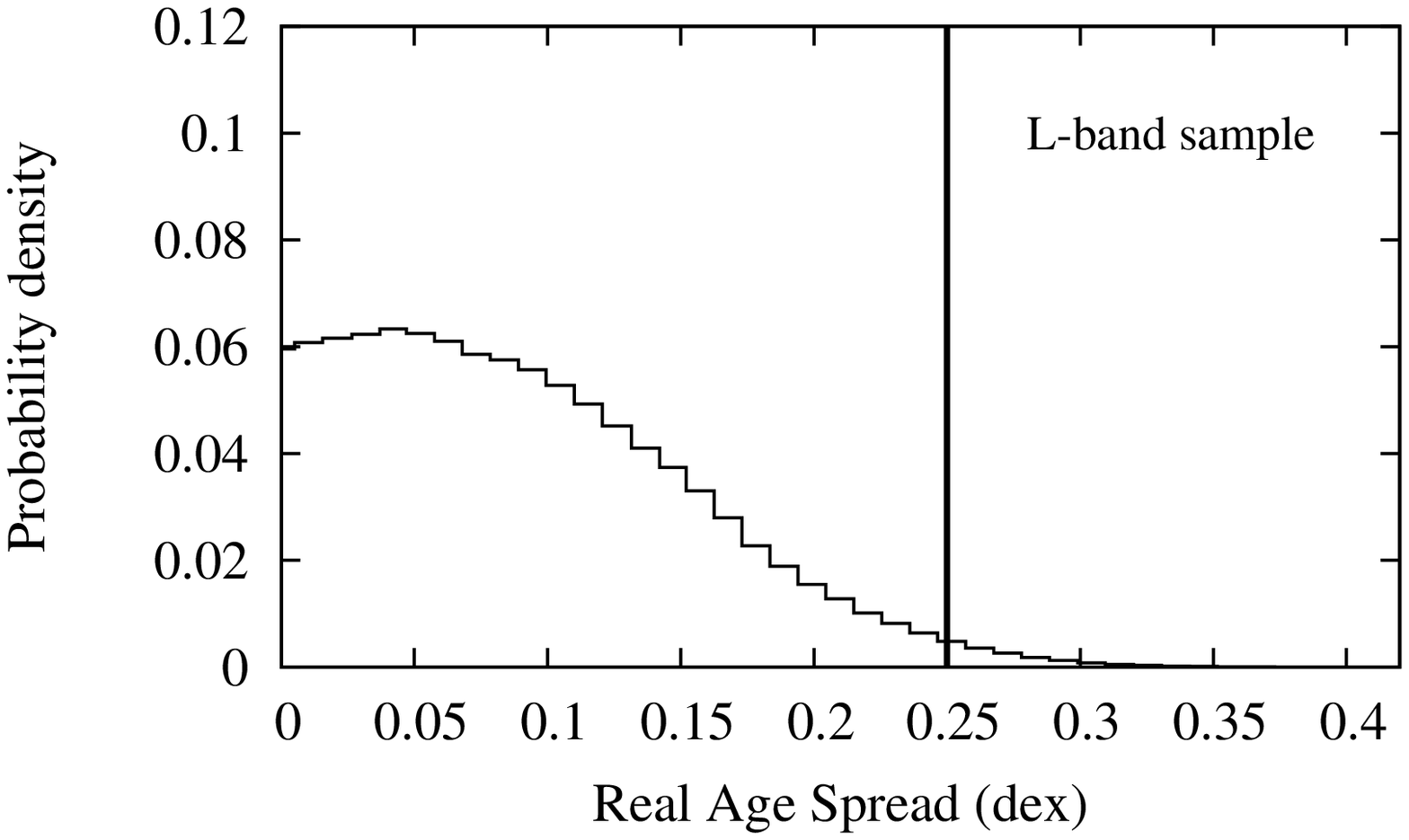}
\includegraphics[width=60mm]{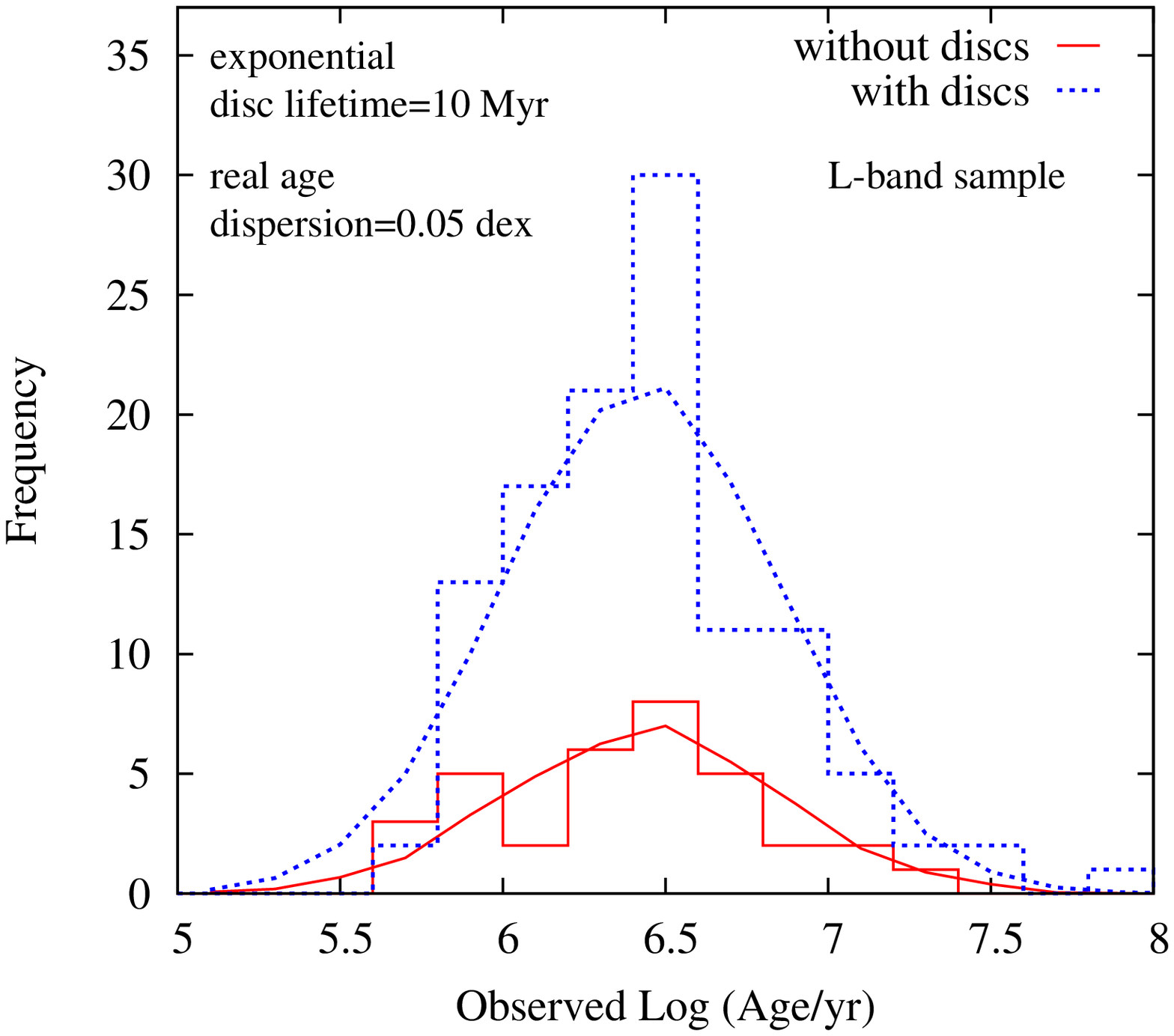}
\end{minipage}
\begin{minipage}[t]{0.33\textwidth}
\includegraphics[width=60mm]{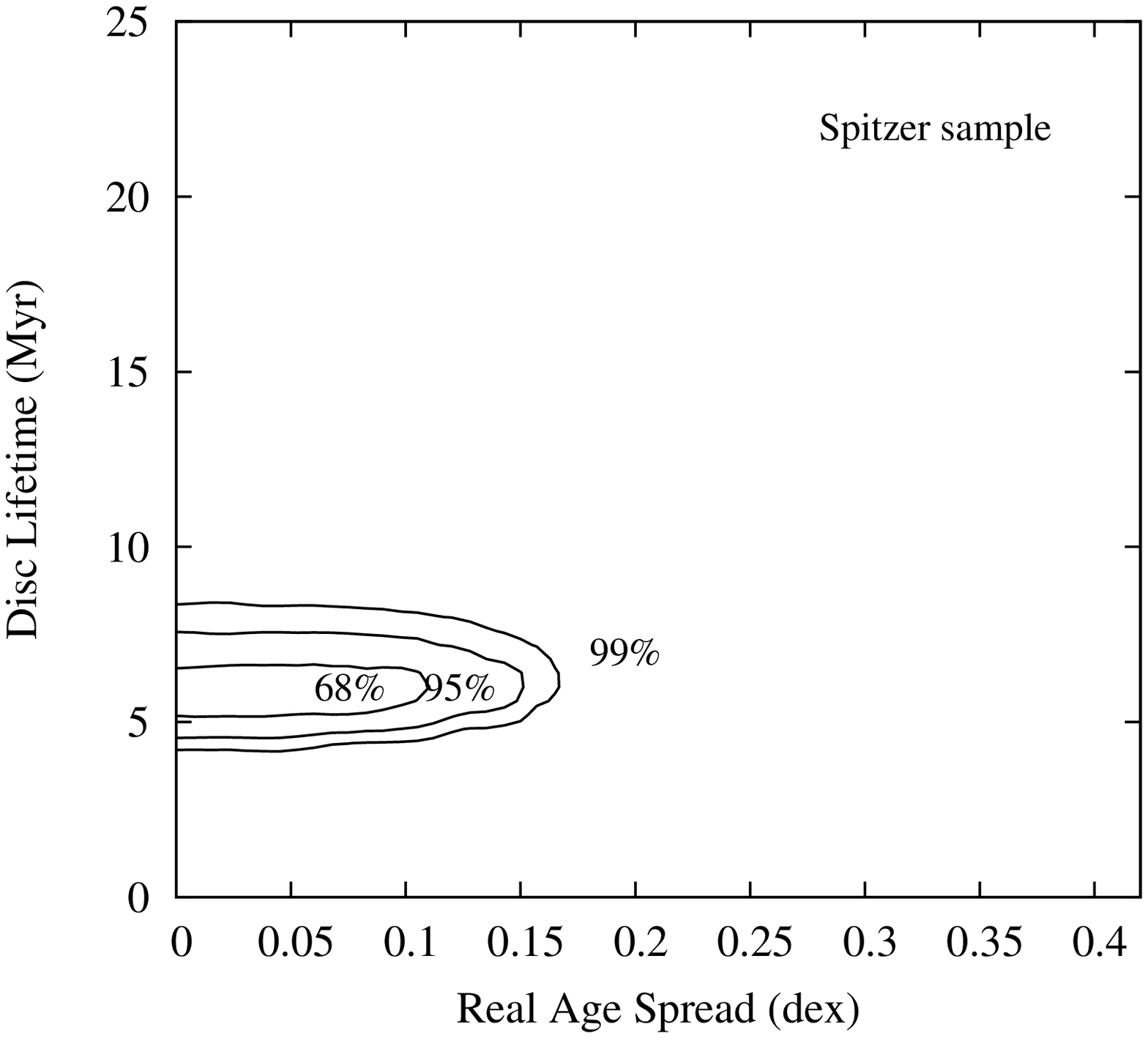}
\includegraphics[width=60mm]{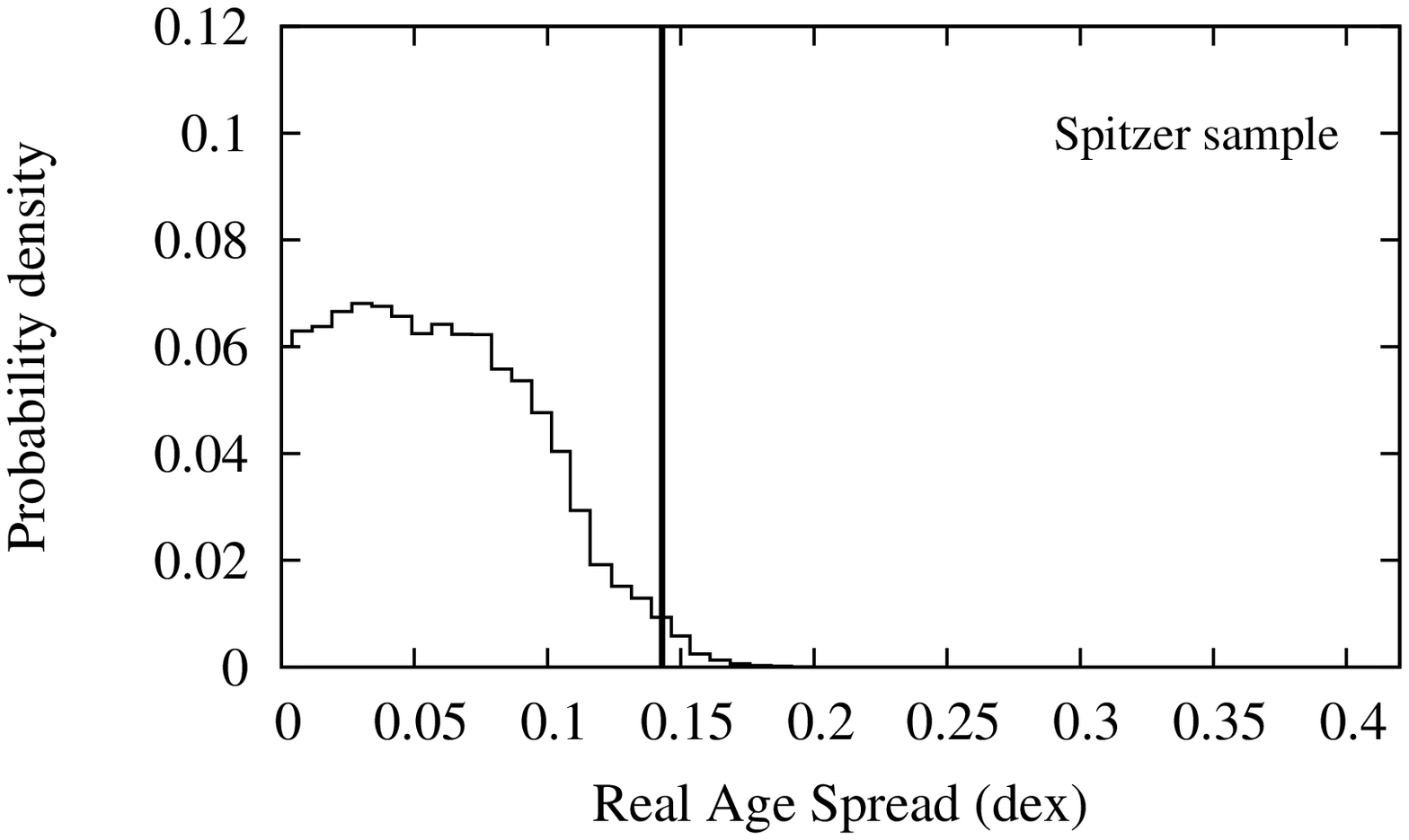}
\includegraphics[width=60mm]{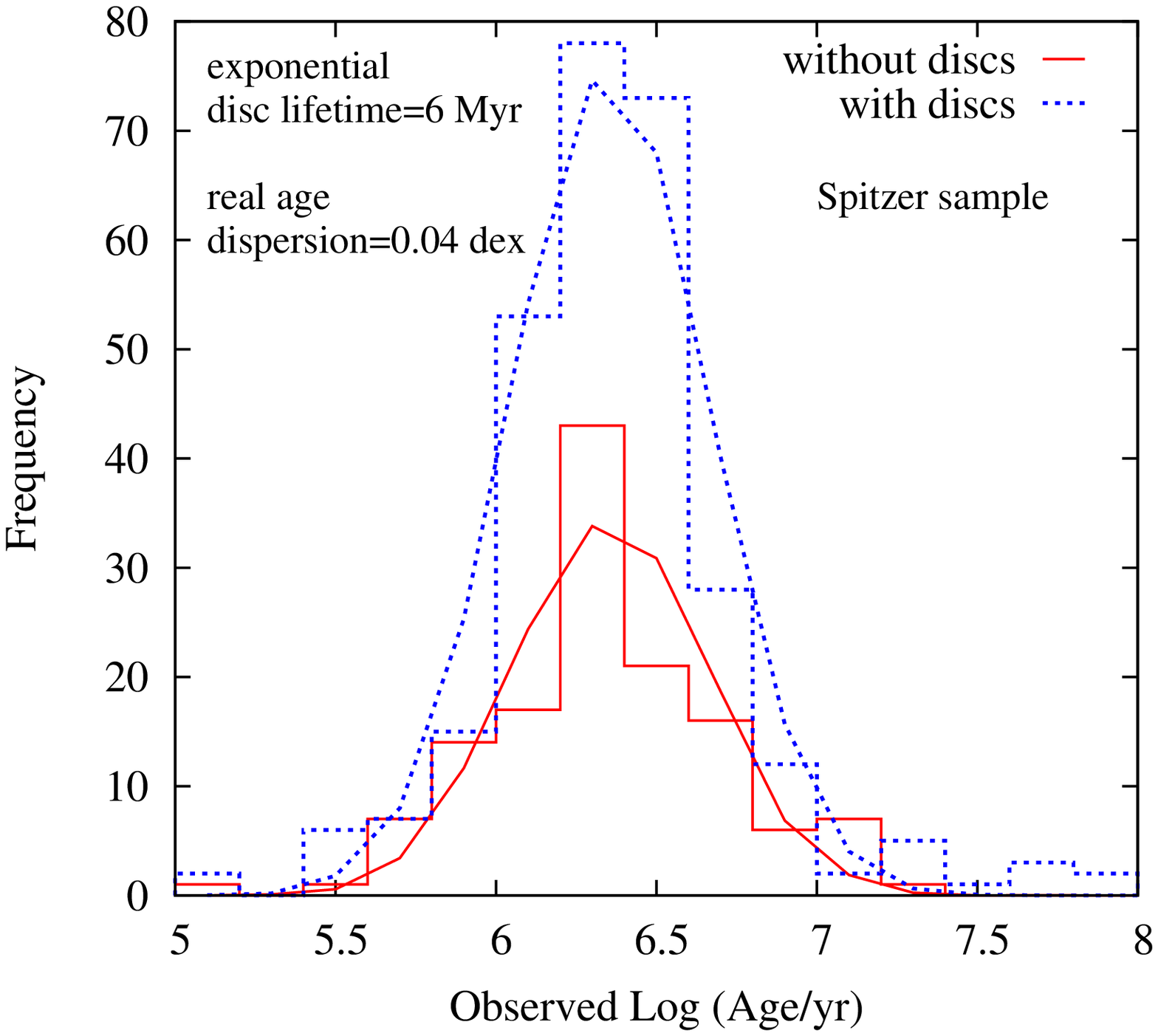}
\end{minipage}

\caption{(Top row): A grid of relative probability for combinations of
  exponential disc lifetime (on the y-axis) and real age dispersion (on
  the x-axis). The contours enclose 68, 95 and 99 per cent of the
  probability. (Middle row): The integral of the probability
  distribution over the full range of possible disc lifetimes, yielding
  the probability distribution for the real age dispersion. 99 per cent
  of the probability distribution lies to the left of the vertical
  lines (note that we integrated over a larger range of disc lifetimes
  than displayed in the top plots).  (Bottom row): The most probable
  age distributions for stars with and without discs. In each row the
  leftmost plot corresponds to discs diagnosed by a $K$-band excess,
  the middle plot corresponds to discs diagnosed by a $L$-band excess
  and the rightmost plot corresponds to discs diagnosed with Spitzer.}
\label{plotmodels}
\end{figure*}

\begin{table}
\caption{Results from modelling the age distributions of stars with and
  without discs and the fraction of stars with discs in the three
  samples discussed in the text. In each case the mean age is set to
  the observed value from Table~1 and the rows list the adopted
  apparent age spread, the derived best-fit real age spreads and
  exponential disc lifetimes and the 99 per cent confidence upper limit to any
  real age spread.}
\begin{tabular}{lccc}
\hline
                & $K$-band &  $L$-band & Spitzer \\
\hline
&&&\\
Apparent age spread & 0.3 & 0.42 & 0.3 \\ 
(dex) &&&\\
&&&\\
Best-fit real age   & $0.09^{+0.02}_{-0.02}$ & $0.05^{+0.12}_{-0.05}$ & $0.04^{+0.06}_{-0.04}$ \\
dispersion (dex) &&&\\
&&&\\
Best-fit exponential &  $4\pm 1$ & $11\pm 3$ & $6\pm 1$ \\
disc lifetime (Myr) &&&\\
&&&\\
99 per cent upper limit & $<0.16$ & $<0.25$ & $<0.14$ \\
to real age spread (dex) &&&\\
&&&\\
\hline
\end{tabular}
\label{modelresults}
\end{table}

Having established that the mean ages and age distributions of stars
with and without discs in the ONC are not significantly different, this
Section develops a simple model to interpret the results quantitatively
and to set limits on the age spread that could be present in the
population and yet still be consistent with the observational data.

\subsection{Model Construction and Parameter Estimation}
The basic model is similar to that described in Section~2.1.  The
distribution of stellar ages, as measured in the HR diagram, is assumed
to be log-normal, with an overall dispersion consisting of the
quadrature sum of two components -- a real age spread and another
dispersion term representing uncertainties in the observations or other
sources of astrophysical scatter in the observed luminosity at a given
age.  The mean age and overall dispersion were chosen to match the
combined apparent age distribution of stars both with and without
discs. The real age dispersion was left as a free parameter.  The
additional observational dispersion was not a free parameter; it was
varied so that when combined with the real age dispersion, the observed
age dispersion was recovered.  The other main component of the model is
a prescription for the disc lifetime. Our most basic assumption is that
most stars begin life with a circumstellar disc that betrays its
existence via the infrared diagnostics we have discussed. The fraction
of stars exhibiting these disc signatures is then assumed to decay
monotonically (on average) with age. Initially we assumed that all PMS
stars begin with a disc and that the disc lifetime obeys a probability
distribution that decays exponentially, with a mass-independent decay
timescale that was a free parameter.

For a given real age spread (characterised by a Gaussian sigma in log
age) and disc decay timescale, the model predicts the {\it apparent}
age distributions of stars with and without discs and the fraction of
stars that still possess discs, within a total population with a given
observed mean age and apparent age dispersion. Separate K-S tests were
performed for the cumulative apparent age distributions of the stars
with and without discs versus their respective model distributions, and
a chi-squared test was performed between the observed and modelled
fraction of stars that still possess a disc.  The product of the three
probabilities arising from these tests was used to estimate the overall
probability that the data were drawn from a population described by the
model. This probability was calculated over a large, two-dimensional
grid of possible values for the real age spread and disc decay
timescale. Examples are shown in Fig.~\ref{probgrids}, calculated by
comparison with the Spitzer sample.

For small sample sizes (the $L$-band sample), the relatively crude
initial assumption of a log-normal age distribution was reasonable and
we found areas of parameter space that gave satisfactory fits to the
data.  The larger $K$-band and Spitzer samples revealed deficiencies in
this simple model -- the observed log-normal age distributions in these
samples have a significant kurtosis of $2.43\pm 0.21$ and $4.84\pm
0.24$ respectively, and are more peaked than a simple Gaussian, with
extended wings. The broad Gaussians required to match the overall age
dispersion are not a good fit near the observed distribution medians
and thus result in very low K-S probabilities.  To counter this, and
provide a slightly more conservative (i.e. larger) upper limit to the
possible real age dispersion, we allowed the input dispersion of the
apparent age distribution to vary from the observed values of 0.38\,dex
and 0.41 dex (see Table~1), finding that a smaller value of 0.3\,dex
gave a much more probable model in both cases. This represented the
core of the apparent age distribution quite well (see bottom row of
Fig.~\ref{plotmodels}). The derived parameters (and limits) are in fact
rather insensitive to this procedure, because of the very low weight
that is attached to objects in the tails of the apparent age
distribution by K-S tests.\footnote{We have experimented with clipping
out 10--20 per cent of the sample as outliers (both in the data and
models) and find almost no quantitative difference from the results in
Section 4.2 where all the data were included.} The nature of these
outliers and whether they offer any support to the idea of a real age
spread is discussed further in Section 5.3.

Confidence intervals on the model parameters were estimated by
renormalising the probability grid so that the sum over all possible
parameter combinations was unity. Contours containing arbitrary
fractions of the probability distribution were calculated from this
grid. Integrating the grid over one or other of the parameter axes gave
estimated confidence intervals in one parameter. The extent of the
probability grids were larger along the disc lifetime axis (typically
up to 100\,Myr) than displayed in Fig.~\ref{plotmodels} to ensure that
all probability was accumulated.

\subsection{Model Results}

The results of comparing the models with the three ONC data samples are
tabulated in Table~\ref{modelresults} and illustrated in
Figs.~\ref{probgrids} and~\ref{plotmodels}. Fig.~\ref{probgrids}
demonstrates to what extent the derived model parameters are sensitive
to each of the observational constraints using the example of the
Spitzer sample. It can be seen that the disc lifetime is very strongly
constrained by the observed fraction of stars with discs. The real age
spread is strongly constrained by the age distribution of stars that
have lost their discs, whereas the age distribution of stars with discs
rules out parameter space featuring large age spreads and short disc
lifetimes.

The outcome is a consistent interpretation from all three samples,
varying in statistical significance as expected from the different data
set sizes.  Figure~\ref{plotmodels} shows that the lack of any
difference in the observed age distributions of the stars with and
without discs constrains the real age spread to be much lower than the
observed age spread and formally consistent with zero for all three
samples. The most constraining dataset is the large Spitzer sample,
which demands that the real age spread be $<0.14$\,dex with 99 per cent
confidence. Even the smaller $L$-band sample provides a 99 per cent
upper limit to the real age spread of $<0.25$\,dex.  The disc lifetime,
as parameterised in this model, is also well constrained in the two
larger datasets at about $4\pm 1$\,Myr ($K$-band sample) or $6\pm
1$\,Myr (Spitzer sample), corresponding to a half-life of 3--4\,Myr.
It is larger for the $L$-band sample at $11\pm 3$\,Myr, due to the
higher $L$-band disc frequency.

\subsection{Sensitivity of the Results to Model Assumptions}

\begin{figure*}
\centering
\begin{minipage}[t]{0.33\textwidth}
\includegraphics[width=60mm]{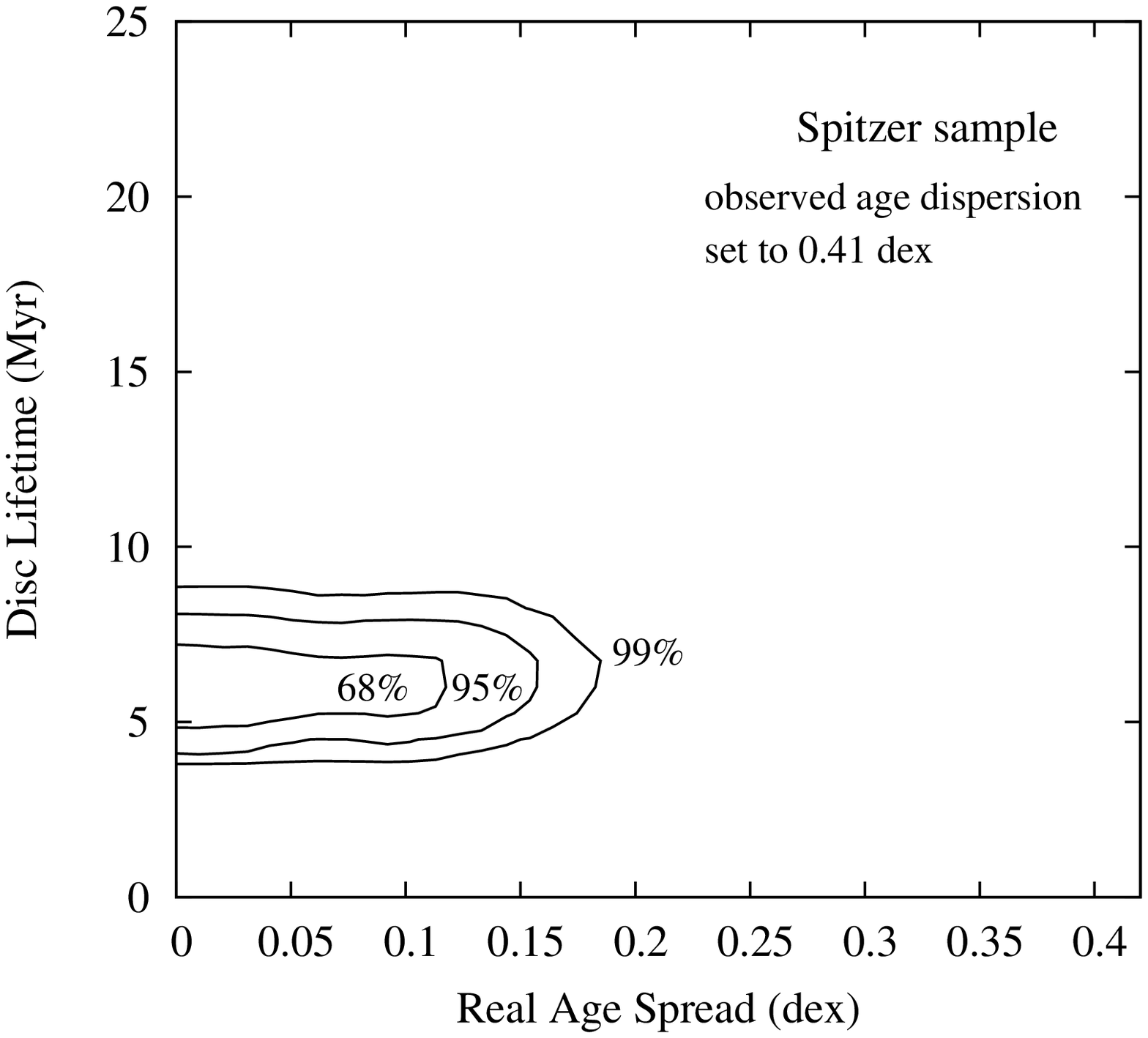}
\includegraphics[width=60mm]{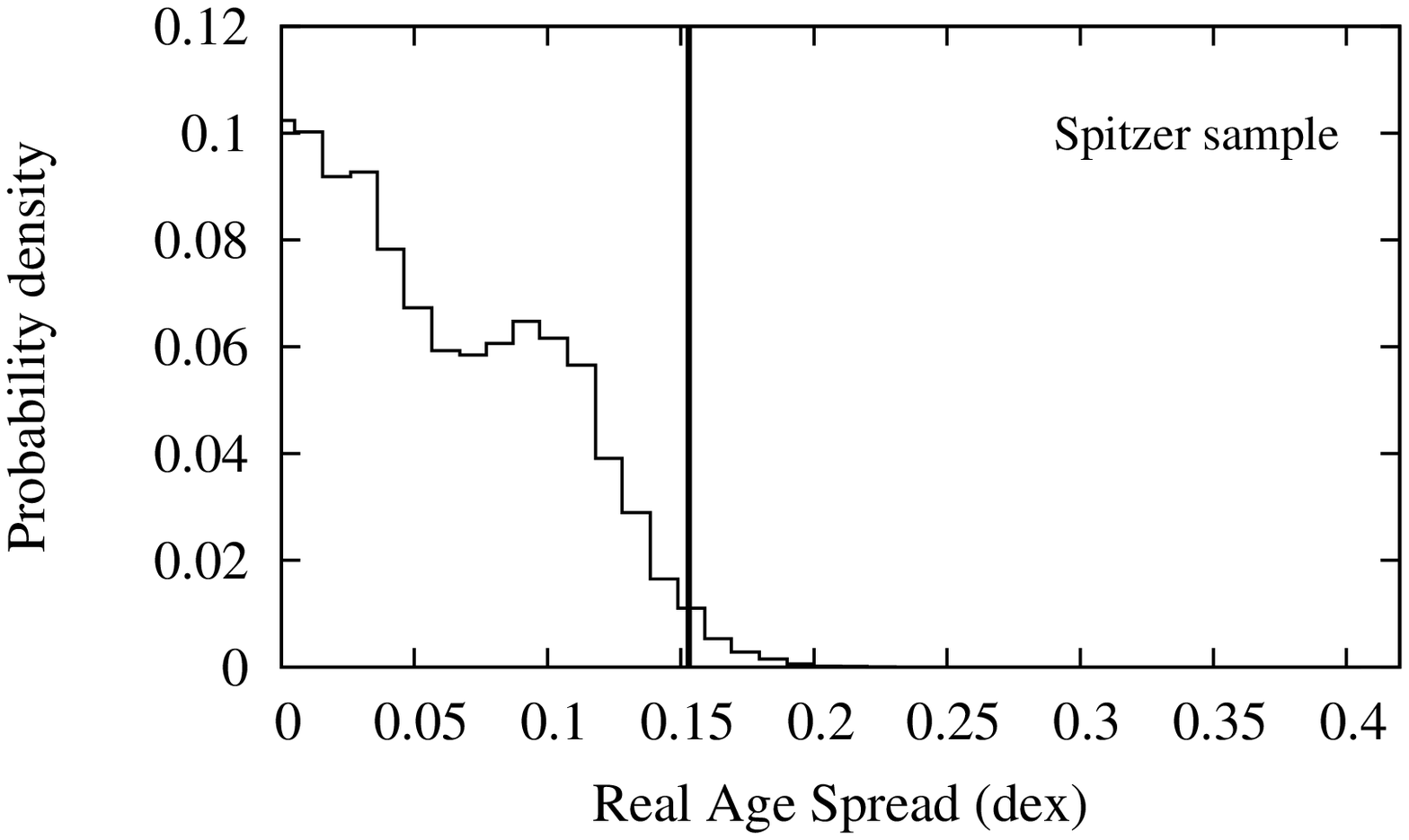}
\includegraphics[width=60mm]{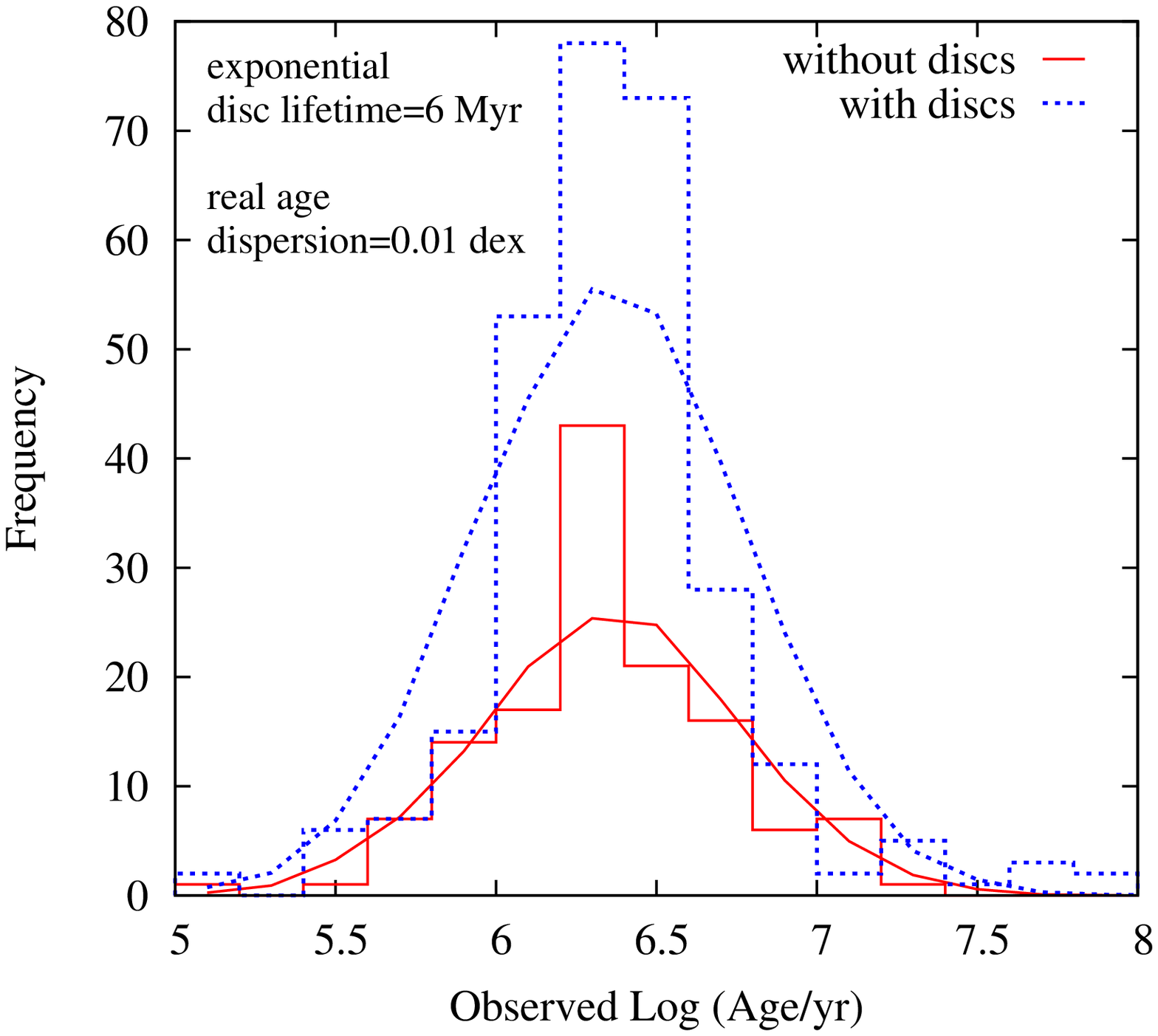}
\end{minipage}
\begin{minipage}[t]{0.33\textwidth}
\includegraphics[width=60mm]{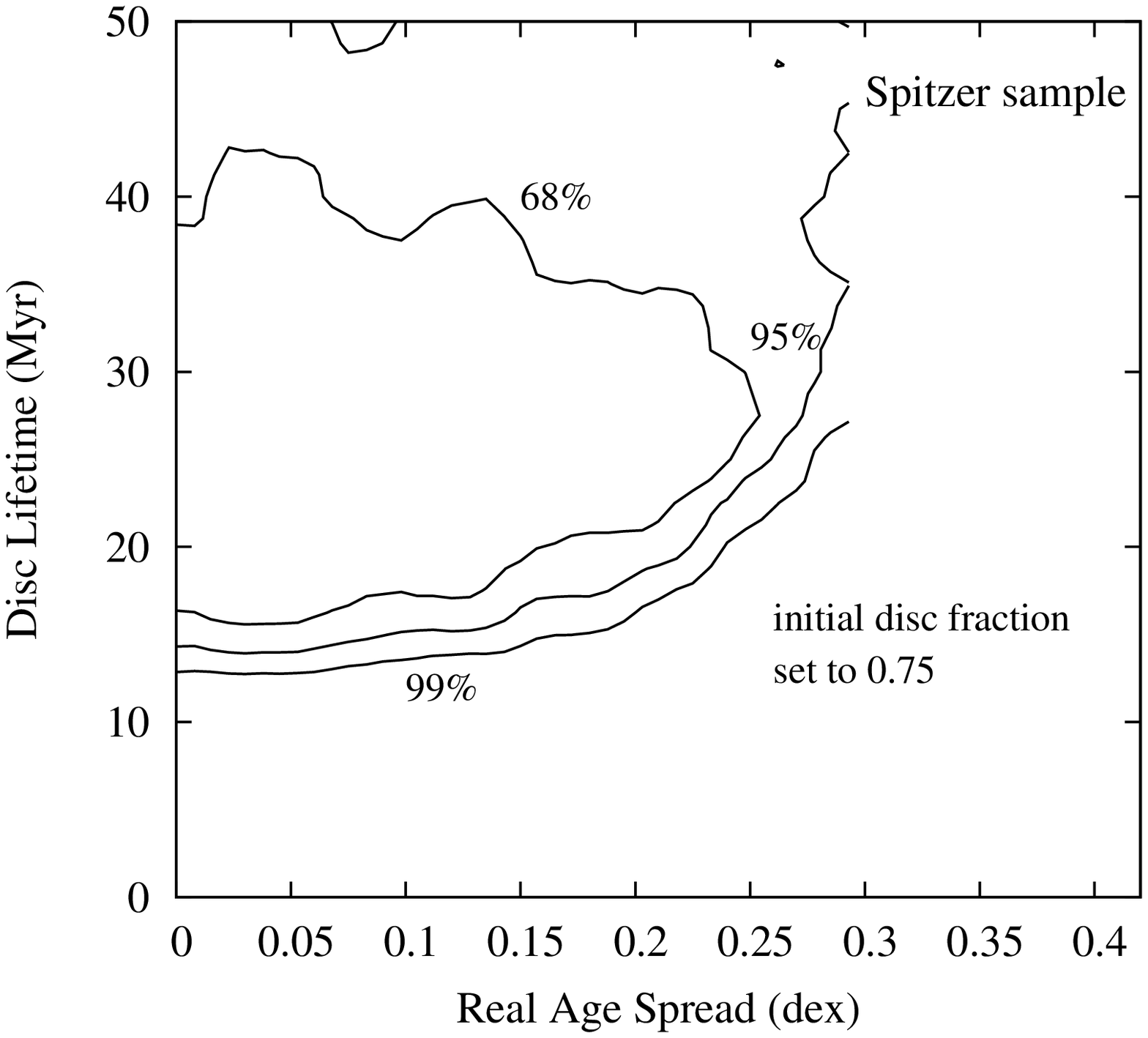}
\includegraphics[width=60mm]{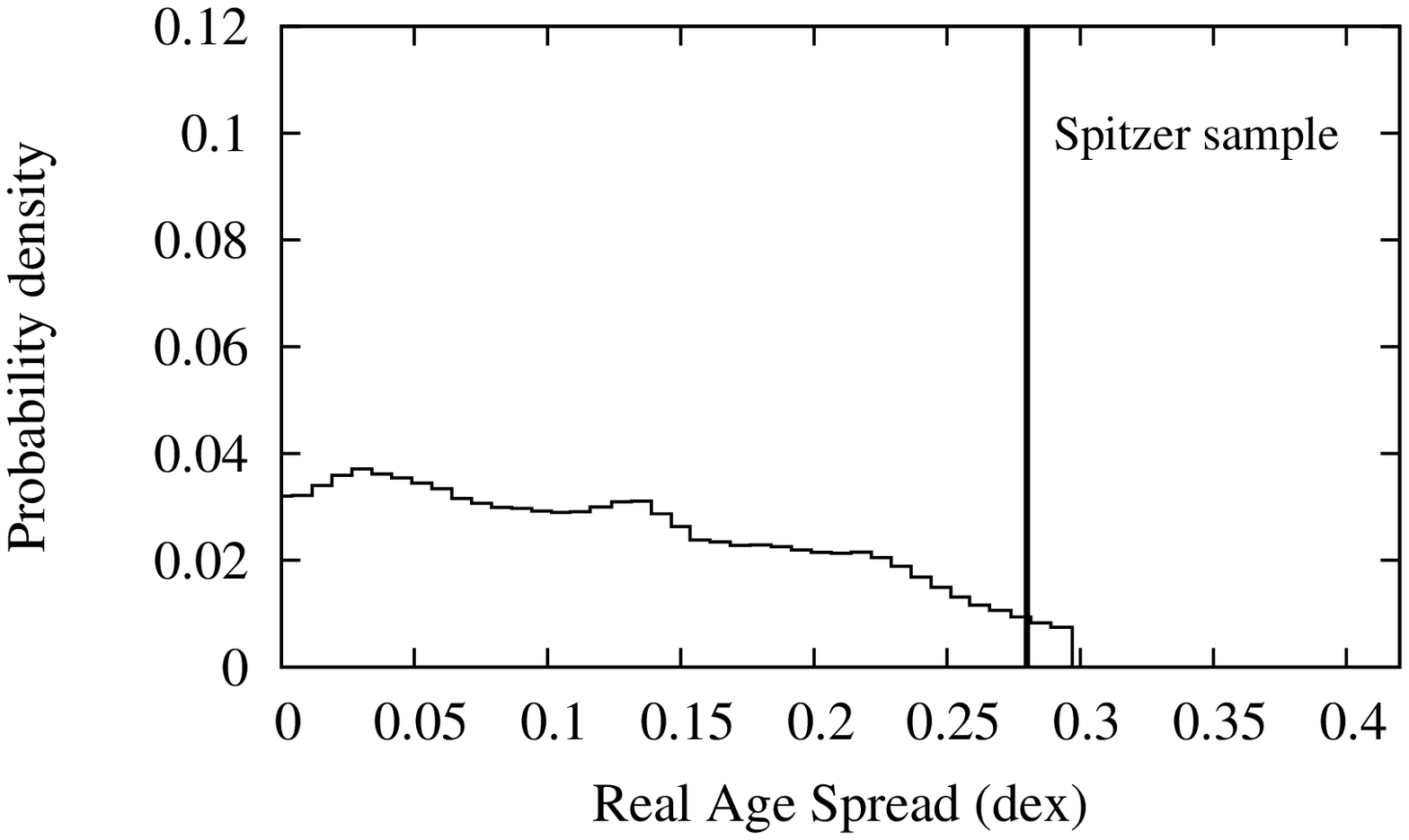}
\includegraphics[width=60mm]{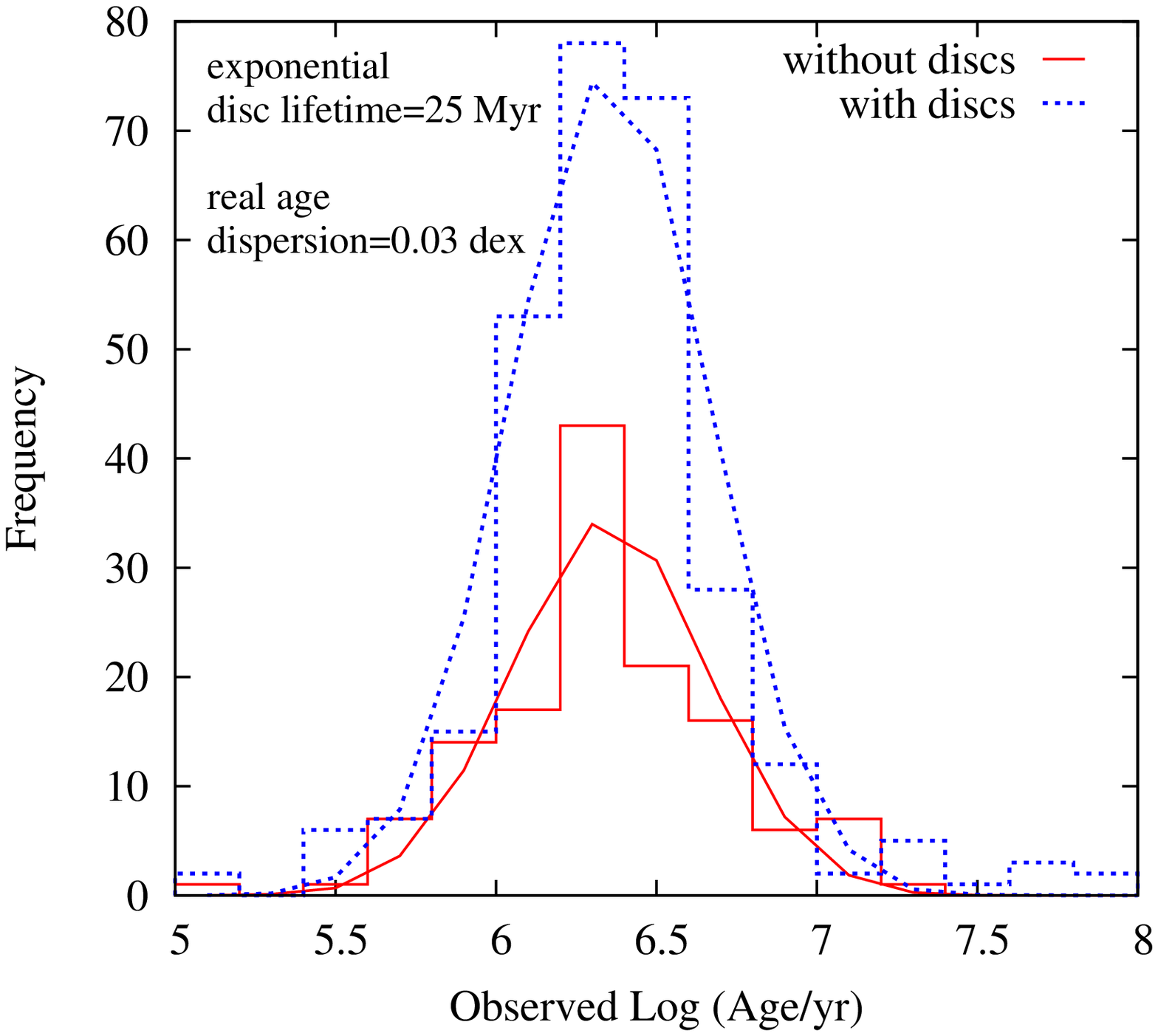}
\end{minipage}
\begin{minipage}[t]{0.33\textwidth}
\includegraphics[width=60mm]{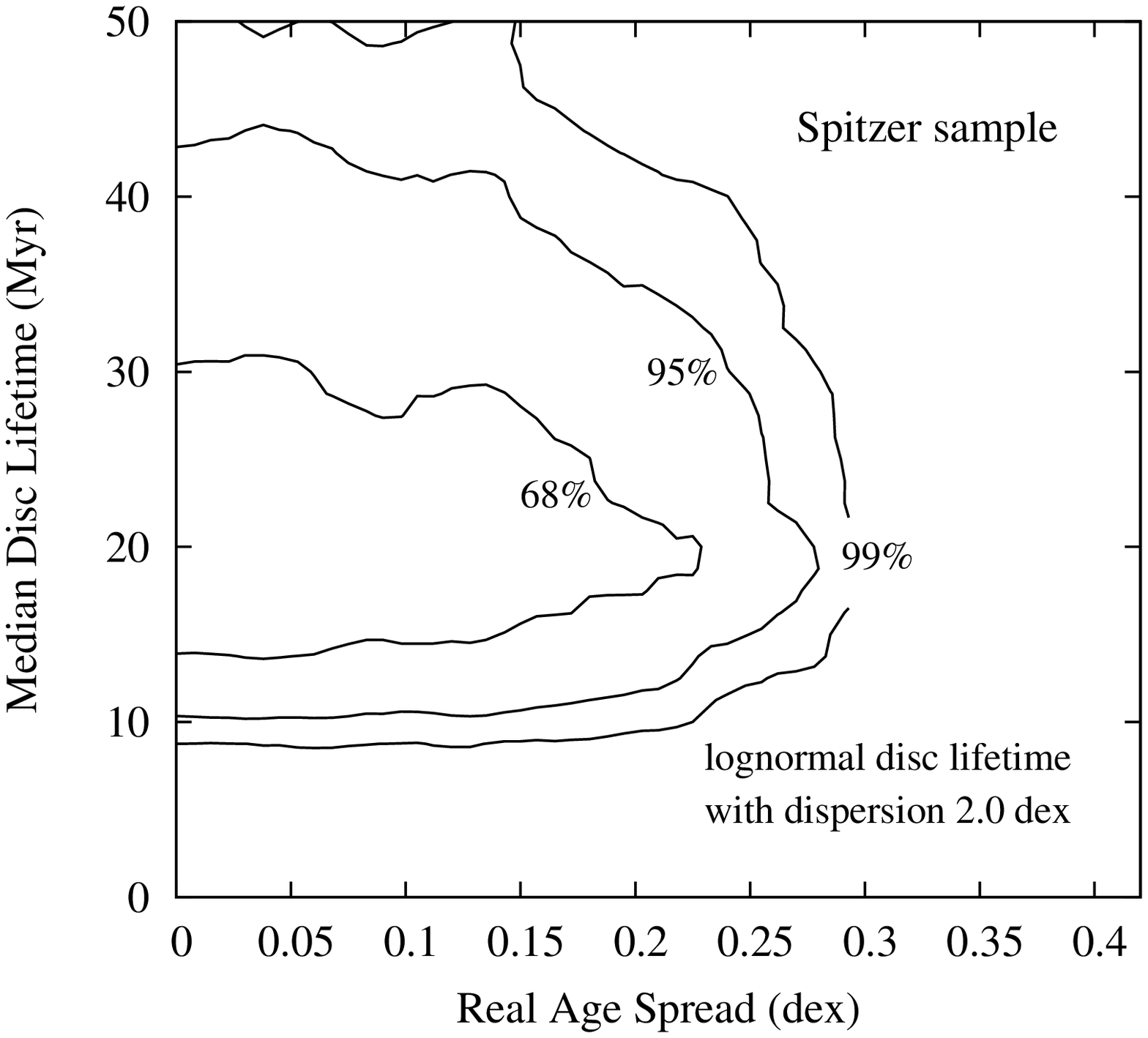}
\includegraphics[width=60mm]{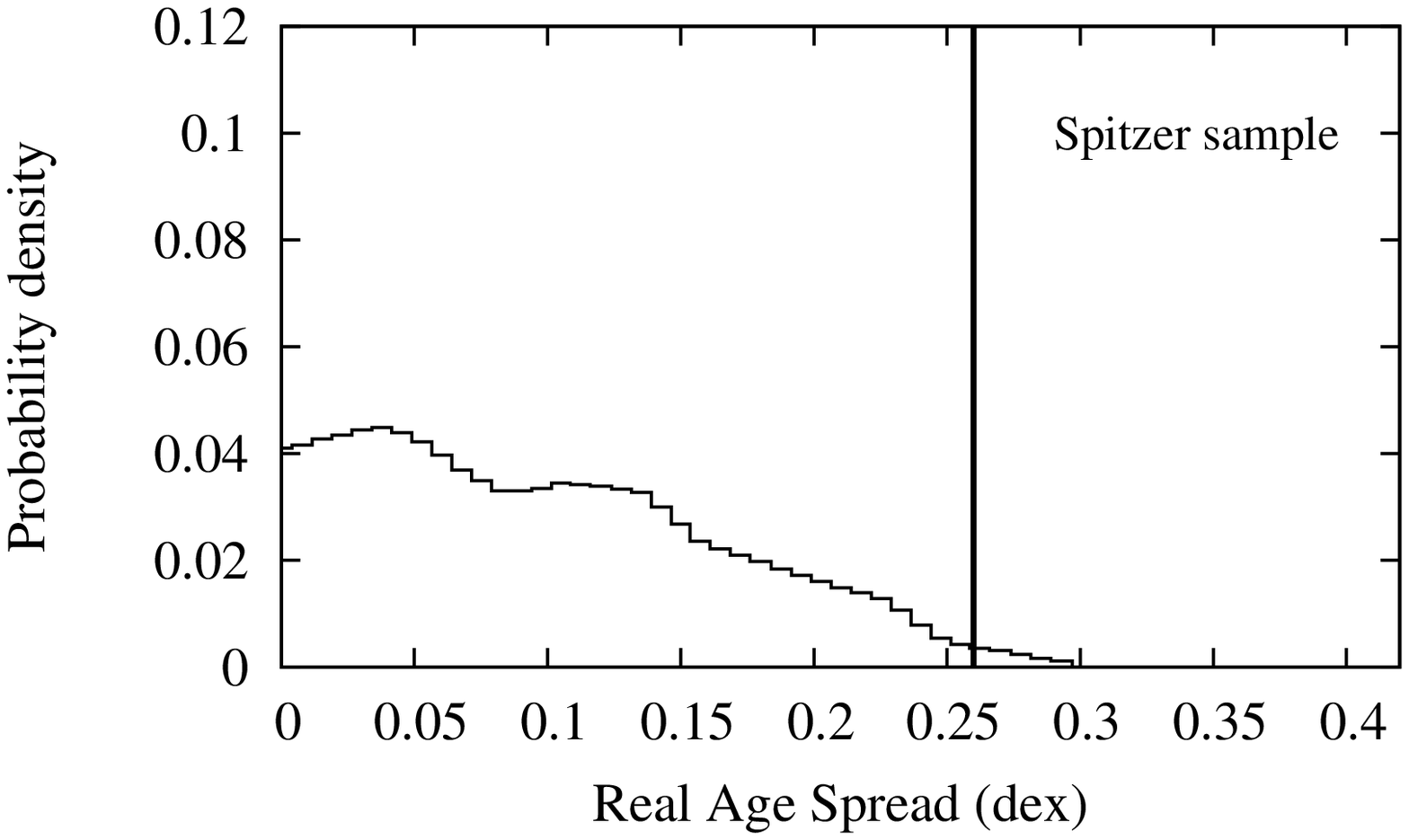}
\includegraphics[width=60mm]{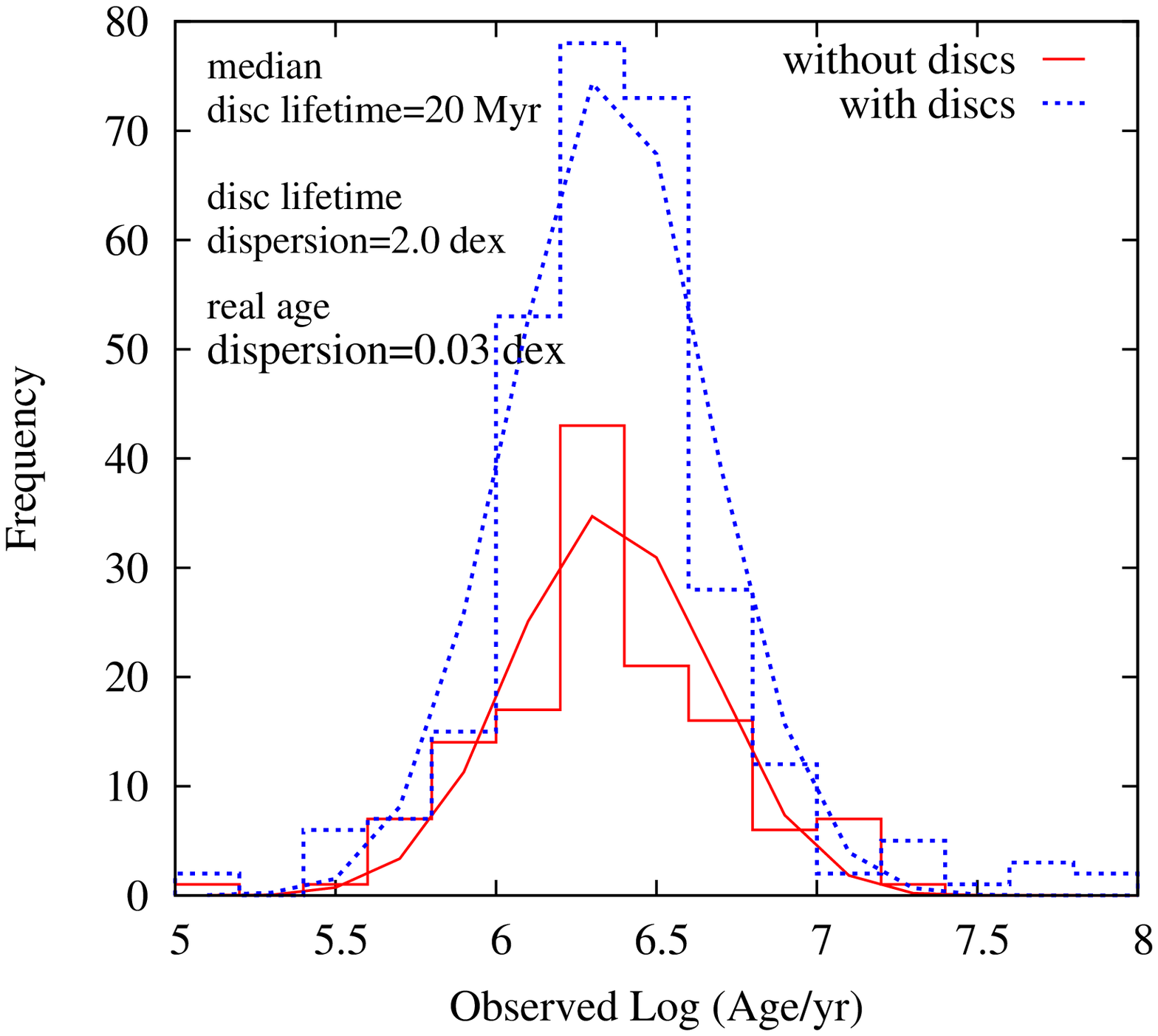}
\end{minipage}

\caption{Exploring sensitivity to key model assumptions.  These plots
  illustrate results for the Spitzer sample and are analogous to the
  third column of Fig.~5.  The results for the $K$-band and $L$-band
  samples are qualitatively similar.  (Column 1): As per Fig.~4, but
  the apparent age dispersion was fixed at the observed value of
  0.41\,dex rather than 0.3\,dex (as in Fig.~5). The best model is a
  poorer fit to the age distributions than those in Fig.~5 (see bottom
  plots), but the upper limit to any real age spread remains similar.
  (Column 2): As per Fig.5, but the initial disc fraction was set to
  0.75 instead of 1.0. This leads to longer inferred disc lifetimes and
  allows a larger real age spread compared to Fig.~5.  (Column 3): As
  per Fig.~4, but the disc lifetime distribution is log-normal in age
  with a dispersion of 2.0\,dex. The y-axis of the top plot now
  corresponds to the median disc lifetime of this distribution. This
  model demands a longer median disc lifetime which then also permits a
  larger real age spread than in Fig.~5, but which is still less than
  the total observed age spread of 0.41\,dex.}
\label{extras}
\end{figure*}

The sensitivity of the conclusions to various
model assumptions has been investigated. In particular,
the possibilities of salvaging a real age spread that is anywhere near
comparable with the observed age spread were explored in detail. 
Figure~\ref{extras} shows
plots equivalent to those in Fig.\ref{plotmodels}, applied to the
Spitzer sample, corresponding to the following model alterations.
\begin{enumerate}
\item
In column 1 of Fig.~\ref{extras} the input value of the
apparent age spread is increased to its observed value of 0.41\,dex. As
discussed in Section 4.1, the model is a poorer fit to the age distributions
of stars with and without discs and the upper
limit to the real age spread is almost unaltered, at $<0.15$\,dex.
\item
In the initial models we assumed all stars started with discs. 
A lower initial disc frequency would imply that even at ``zero age'' there are some stars without
discs. This has the effect of allowing much longer disc
lifetimes, because fewer stars need to lose their discs to explain
the observed disc frequency at a later time, and it simultaneously relaxes the constraint
on any real age spread because even quite old stars could have a
similar probability to young stars of being discless.
Presumably, the initial disc fraction cannot be
lower than the currently observed disc fractions.
Considering the Spitzer sample with an observed disc frequency of
0.68, then if we adopt an initial disc
fraction of 0.75, this allows a 99 per cent upper limit to the real age
spread of 0.28 dex, which is an appreciable proportion of the
observed age spread, although the best-fitting model still has a very
small real age dispersion. This is achieved at the expense of a very long
disc lifetime of $>20$\,Myr (see column 2 of Fig.~\ref{extras}).
\item 
A different functional form for the disc lifetime distribution may also
allow longer disc life times.  Rather than an exponential decay, the
disc lifetimes could be distributed normally in log age in a similar
way to the observed ages of the stars. If the dispersion of this normal
distribution is smaller than the observed age spread (i.e. $<
0.4$\,dex), then the results are very similar to those shown in
Fig.~\ref{plotmodels}. However, if the dispersion in disc lifetimes is
made much wider around some median value, then there can be a significant
probability that an older star would have kept its disc and likewise
that a younger star had lost its disc. This means that the apparent age
distributions of stars with and without discs could be quite similar
even in the presence of a real age spread. Column~3 of
Fig.~\ref{extras} shows an extreme example where we have allowed the
dispersion in disc lifetimes to be as large as 2.0 dex (almost a flat
distribution over the entire range of observed ages). This huge range of disc
lifetimes is very unlikely from other considerations (see section 5), but it
could simultaneously explain the presence of ``young'' stars without
discs and ``old'' stars with discs in the case of a real age spread as
large as $\simeq 0.2$\,dex, as long as the {\it median} disc lifetime is
$>10$\,Myr.
\end{enumerate}

\section{Discussion}

Assuming that most stars are born with a disc and that the frequency of
stars exhibiting disc signatures decays (on average) monotonically,
then the similar apparent age distributions of stars with and without
discs in the ONC imply that there is unlikely to be a large real age
spread within the bulk of the population. The simple quantitative model
developed in Section~4 suggests that any real age dispersion is limited
to $<0.14$ dex at 99 per cent confidence -- a small fraction of the
observed $\sim 0.4$ dex apparent age dispersion. This additional
dispersion would therefore have to be due to observational
uncertainties or other sources of astrophysical scatter in the stellar
luminosities that are not related to age.

Our simple model could be criticised in that it does not provide exact
fits to the observed age distributions, and the approach taken to
modelling disc lifetimes is simplistic. In Section 4.3 we considered
alternative disc lifetime distributions that do allow a larger fraction
of the observed age spread to be real. These involve either initial
disc fractions smaller than unity or a very large spread of disc
lifetimes. Either is perhaps possible for the ONC, but the currently
observed disc frequency would then require that median disc lifetimes
be very long ($>10$ Myr). A more robust qualitative statement of our
conclusion for the ONC is therefore that any real age spread in the PMS
population is smaller than the median disc lifetime, rather than
smaller than some absolute value. The possibility of long disc
lifetimes, sufficient to allow a large real age dispersion, is
permitted by the ONC data alone, but unlikely in the context of
infrared observations of other clusters. A median disc lifetime
$>10$\,Myr would mean that a large fraction of stars in clusters with
ages $\simeq 10$\,Myr would still possess inner discs, but that is not
borne out by the observational facts. Older clusters have infrared disc
fractions limited to at most a few per cent \citep[e.g.][]{sicilia06,
hernandez08}.

\subsection{Selection Effects and Biases}

Incompleteness could perhaps explain similar age distributions for
stars with and without discs if our samples were missing either
low-luminosity stars without discs or stars with discs among the more
luminous, apparently young stars. The parent sample from DR10 does
become increasingly incomplete for less luminous ``older'' stars, but
this is not an issue if it applies equally to stars with and without
discs. The infrared samples are also incomplete for less luminous
objects, but this incompleteness will be less severe for stars with an
infrared excess, because by definition these are brighter in the
infrared. In principle then, ``older'' stars with an infrared excess
are more likely to appear in these catalogues than stars of a similar
age without discs. The HR diagrams in Fig.~3 allow the reader to assess
how problematic this may be. It is clear that the infrared data are
sampling stars from almost the full range of the HR diagram of the
parent sample, although there is some evidence that fainter stars lying
below the 10\,Myr isochrone in the Spitzer sample are more likely to be
stars with discs. This could be due to the effect we are discussing,
but it could also arise from the possibility that some of these very
low-luminosity sources are occulted by edge-on discs and being observed
via scattered light (see Section 5.3). We do not expect either of
these possibilities to significantly affect our results because the K-S
tests used here are insensitive to differences in the tails of the
respective apparent age distributions. In any case, the $L$-band sample
is almost complete (in terms of having measurements for sources in the
DR10 catalogue in the same area of sky -- see Fig.~3) and whilst of
lower statistical quality, the results from this sample do not
contradict those from the Spitzer sample.

A less obvious bias might arise if incompleteness in the infrared
samples affects the measured disc frequency. The $K$-band census of
discs is likely to be incomplete towards lower masses - indeed the
$K$-band sample disc frequency is significantly lower than the other
two samples considered. A lower disc frequency leads to a smaller
inferred disc dispersal timescale, which in turn means that a given age
spread should produce a more marked difference in the age distributions
of stars with and without discs. Thus if either the census of stars
with discs is incomplete, or the subsample of stars without discs
contained many contaminating non-members, then a misleadingly low disc
frequency would be measured, the inferred disc lifetime would be too
short, and the real age spread constrained too tightly. This bias is
unlikely to be important here because stars with discs are actually
more likely to be included in the infrared surveys; the samples were
screened to exclude objects with low proper motion membership
probabilities; and in any case if disc frequencies were higher,
allowing long disc lifetimes, then this would contradict the mean disc
lifetime estimates from the ensemble of young cluster observations.

\subsection{What Causes the Apparent Age Spread?}

If the apparent age spread in the HR diagram is not genuinely caused by
a large age dispersion then how does it arise?  Estimates of the
uncertainties in measured luminosities have been made by a number of
authors \citep{hillenbrand97, hartmann01, rebull04, burningham05} and
are likely to have pseudo-Gaussian dispersions in the range
0.16--0.20\,dex. The well-known $L \propto t^{-2/3}$ scaling between
age and luminosity on the PMS would then lead to an estimated Gaussian
dispersion in the apparent age of 0.24-0.3\,dex. Most recently,
\citet{reggiani11} have shown, specifically for the ONC, that
observational uncertainties and variability are unlikely to cause an
apparent age dispersion beyond 0.15\,dex. Thus observational
uncertainties could be a component of the observed apparent age
dispersion, but are unlikely to account for all of it.  Supporting
evidence for this point of view comes from the observed projected radii
of PMS stars in the ONC \citep{jeffries07b}. These exhibit a larger
dispersion than can be explained by random inclination angles, implying
a spread in stellar radii at a given $T_{\rm eff}$ that is consistent
with the observed luminosity spread.  In other words, the dispersion in
luminosity appears to be genuine, with observational uncertainties
playing only a minor role.

A further argument in favour of genuine dispersions in luminosity and
radius, but not age, arises from the work of \citet{littlefair11}.
Here it was found that the rotation rates of high luminosity,
apparently young, stars in the ONC were faster than for the
low-luminosity, apparently old, stars. This contradicts the idea that
the apparently older stars have evolved with time from the positions in
the HR diagram currently occupied by the apparently younger stars,
since in so doing they should have contracted and hence spun up.
Instead \citet{littlefair11} suggest that both groups have a similar
age but that their luminosities, radii and rotation rates have been
affected by prior episodes of heavy accretion.

How could there be a genuine dispersion in luminosity, but not in age?
The idea that early accretion could alter the position of a PMS star in
the HR diagram making it appear younger or older has existed for some
time \citep{mercersmith84, tout99}.  This hypothesis has been revived
by observations of class I young stellar objects
\citep[e.g. by][]{enoch09} that suggest the main mass-building phase of
a star could be characterised by transient or episodic accretion at
very high rates ($\sim 10^{-4}\,M_{\odot}$\,yr$^{-1}$) for brief
periods of time ($\sim 100$\,yr). This episodic accretion has been
modelled by \citet{vorobyov06} and its consequences for PMS
evolutionary tracks were explored by \citet{baraffe09}.  They find that
if accreted energy can be efficiently radiated away, then a short phase
of rapid accretion builds up the stellar mass, but leaves insufficient
time for the star to adjust its radius before it emerges on the PMS. In
the Baraffe et al. (2009) models, the consequent PMS star emerges on
the HR diagram after the class I phase with a smaller radius and lower
luminosity than would be otherwise predicted by non-accreting
evolutionary tracks and {\it appears} older.  A range of accretion
histories could lead to a dispersion in the observed luminosities among
any coeval group of stars with ages less than the Kelvin-Helmholtz
timescale of $\sim 10$\,Myr.  As there may be no connection between
accretion rates in the class I phase and later accretion as a class II
T-Tauri star, this would effectively randomise the apparent ages
determined from the HR diagram for young PMS stars.

Other authors suggest that the effects of early phases of heavy
accretion may not be so dramatic. The models of \citet{hosokawa11} show
that non-accreting isochrones may only overestimate true ages for PMS stars
with $T_{\rm eff}>3500$\,K (about half of our sample). \citet{hartmann11}
argues that the early accretion phase probably adds significant energy
to the protostar, perhaps increasing rather than decreasing the
radius. Nevertheless they also argue that plausible variations in
initial protostellar radii and accretion histories could give rise to
0.3\,dex dispersions in apparent stellar age -- a significant
proportion of that observed.

\subsection{The Consequences for Ages, Age Spreads and Cluster
  Formation Timescales}

If our approach and assumptions are valid then ages from the HR diagram
cannot be used reliably to trace the history of star formation in the
ONC as attempted by \citet{palla99} and \citet{huff06}. Furthermore,
our work implies that the bulk of the stars in the ONC are formed over
a timescale shorter than the median lifetime of circumstellar
material. Our basic quantitative model suggests that if the ONC has a
mean age of about 2.5\,Myr (see Table~1), then at least 95\% of its
stars must have ages of 1.3--4.8\,Myr (based on a 1-sigma dispersion of
0.14\,dex and a log normal age distribution) with a more likely range
that is smaller than this. Note though that this mean age, and hence
age range, are dependent on the adopted evolutionary models, which have
significant systematic uncertainties at these young ages
\citep{baraffe02}.

The ONC age has been previously estimated at 2--3\,Myr using PMS
isochrones and the recently revised ONC distance of $\simeq 400$\,pc
\citep{jeffries07a, sandstrom07, menten07, mayne08}. DR10 obtained an
ONC age of 2.6\,Myr based the isochrones of \citet{siess00}.
\citet{naylor09} provides a uniform recalibration of young PMS ages
based on fitting of the upper main-sequence, finding an age for the ONC
of 2.8--5.2\,Myr and there is a kinematic age of $\geq 2.5$\,Myr, found
by tracing back three runaway stars to their estimated origin as a
single stellar system in the ONC \citep{hoogerwerf01}. On the other
hand a younger mean age would agree better with the conclusions of
\citet{tobin09}, who argued that the close similarity between the
kinematic structure of the stars and gas in the ONC indicate that the
cluster is no more than a crossing time old \citep[$\sim 1$\,Myr -- see
also][]{proszkow09}. A younger age might also be indicated by the young
($\la 1$\,Myr) age deduced for the high-mass stars in the Trapezium
\citep{storzer99, clarke07, mann09} Our analysis could be consistent
with the adoption of any of these estimates for the mean ONC age
because we only require that any age spread is smaller than the
disc dispersal timescale.

Whilst the ages of individual young PMS stars derived from the HR
diagram may be unreliable, this does not necessarily invalidate mean
ages deduced for whole clusters, because the additional sources of
luminosity scatter may be roughly symmetric. Even if they were not,
there is no reason to suppose that the rank ordering by mean age of
well-observed nearby clusters is incorrect and so there is no
contradiction in our use of mean cluster ages to argue for a monotonic
decay of disc indicators with age (although the absolute timescale must
be uncertain), whilst at the same time arguing that individual stellar
ages are so uncertain as to preclude using them to estimate star
formation histories.

\citet{huff06} suggest that star formation began in the ONC at a low
level as long ago as 10\,Myr and indeed there are stars in the HR
diagram that are as old or older than this. However, as our modelling
is insensitive to the tails of the age distributions (see Section 4.1),
it is reasonable to ask whether a small population of older stars might
still be consistent with our analysis. But there is a problem with this
idea when confronted with the infrared disc diagnostics, because too
many of these ``old'' stars still possess discs. For instance, in the
Spitzer sample there are 56 stars (13\% of the sample) with an apparent
age $>5$\,Myr and 36 of them have discs based on their [3.6]-[8.0]
colour, a fraction consistent with the overall disc fraction of
68\%. If the ages of these ``old'' stars were accurate, then for a
reasonable exponential disc lifetime of say 6\,Myr (see Section 4.1),
we would only expect to see 13 stars with discs -- inconsistent with
the observed value at very high significance. Some of the more extreme
objects (in terms of their apparent age) could be examples of stars
occulted by their discs and observed in scattered light
\citep{slesnick04}. \citet{kraus09} find that the components of
(presumably coeval) PMS binary systems frequently exhibit apparent age
differences of 0.4 dex and more, which they also attribute to
systematic problems in estimating the luminosities of PMS stars with
discs and accretion. A similar argument applies to the 56 stars in the
Spitzer sample that are apparently younger than 1\,Myr. Only 32 of
these possess discs compared to the 51 expected for a disc lifetime of
6\,Myr and an initial disc fraction of unity.  In summary, the disc
frequencies also argue against the accuracy of the ages of objects in
both the young and old tails of the age distribution and, like the bulk
of the population, their luminosities must too be explained in terms of
observational uncertanties or physical effects that alter the
luminosity-age relationship.

\section{Summary}

Assuming that most stars are born with circumstellar material and that
the infrared signatures of this material decay, on average,
monotonically with time, then a wide spread of ages ($\simeq 10$\,Myr)
in the ONC should manifest itself by marked differences in the age
distributions of stars with and without infrared disc signatures. This
hypothesis has been tested using a large, homogeneous sample from the
ONC catalogue of DR10, using three independent means of diagnosing disc
presence and ages derived from the HR diagram. We have found no
significant evidence for differences in the apparent age distributions
of stars with and without discs and their means and medians are very
similar.  This is consistent with the conclusion that any real age
spread in the ONC must be smaller than the median lifetime of the
circumstellar discs.

A simple quantitative model has been developed to interpret these
results. This model suggests that for plausible disc lifetimes, the
contribution of any real age spread to the apparent age dispersion
inferred from the HR diagram must indeed be very small; $<0.14$\,dex
dispersion in a log-normal age distribution at 99 per cent confidence,
compared with an observed age dispersion of $\simeq 0.4$\,dex.  Even
stars in the tails of the apparent age distribution
have disc frequencies incompatible with their apparent ages
and consistent with coevality with the rest of the population. These
results argue strongly against cluster formation timescales longer than
a few Myr. If the mean cluster age were $\simeq 2.5$\,Myr, then $>95$\%
of the population has ages between 1.3--4.8\,Myr.

Instead, we suggest that the observed luminosity dispersion in the HR
diagram might be explained in terms of a combination of observational
uncertanties (binarity, extinction, variability etc.) or physical
effects that disorder any simple relationship between luminosity and
age during the early PMS. There is some evidence in the
literature that observational uncertainties alone cannot explain the
full extent of the apparent age dispersion, but regardless of which
mechanism proves to be more important, it may be difficult to use the
presently determined individual ages of PMS stars from the HR diagram
to infer a star formation history for the ONC.  If it does emerge that
physical effects such as the early accretion history of a PMS star can
significantly alter its position in the HR diagram, then even the
average age of the cluster may be unreliable.

It is important to extend the type of analysis described here to other
clusters, using homogeneous techniques to determine luminosity and
$T_{\rm eff}$ and to diagnose the presence of discs, although few
nearby clusters are rich enough to offer the clear statistical tests
provided by the ONC.

\section*{Acknowledgements}
RDJ and NJM acknowledge the support of the Science and Technology
Facilities Council.  SPL is supported by a Research Councils UK
Academic Fellowship.  This research has made use of NASA's Astrophysics
Data System Bibliographic Services. We thank Tom Megeath for supplying
tables of the ONC Spitzer data and a referee for several useful
suggestions.

\bibliographystyle{mn2e} 
\bibliography{iau_journals,master}


\bsp 

\label{lastpage}

\end{document}